\documentclass[12pt]{book}
\usepackage{plenum}
\usepackage{epsf}
\begin{document}

\chapter{
CRITICAL FLUCTUATIONS IN NORMAL-TO-SUPERCONDUCTING TRANSITION
}

\author{
R. Folk,\refnote{1} Yu. Holovatch\refnote{2}}

\affiliation{
\affnote{1}Institut f\"ur Theoretische Physik,
Johannes Kepler Universit\"{a}t Linz
A-4040 Linz, Austria\\
\affnote{2}Institute for Condensed Matter Physics
of the Ukrainian Academy of Sciences,
UA--290011 Lviv, Ukraine
}

\section{1 INTRODUCTION \label{I}}

Recent advances in our understanding of critical phenomena
due to the application\refnote{\cite{Wilson71}}
of the renormalization group (RG) approach\refnote{\cite{Bogoliubov59}}
are now well known (see e.g. the
text\-books\refnote{\cite{Amit84,LeBellac91,Zinn96}}).
The scale invariance at the critical point and the universality of certain
features of  critical
phenomena can be explained by an RG transformation, and lead to a theory
which provides  a quantitative
description of the critical behaviour of various thermodynamic
quantities of interest.

Specifically, the critical point in the RG "language"(in the
context of our lecture we specify it as an equilibrium second order
phase transition point) corresponds to the stable
fixed point of the RG transformation, where the system is scale
invariant. Asymptotic properties of the system are governed solely by
the coordinate of the stable fixed point whereas non-asymptotic ones
are defined in the region of the approach to the fixed point. In addition,
the RG approach may permit the order of the phase
transition occurring in a certain model to be determined. This is explored
by studying the stability of the fixed points of the RG transformation: an
absence  of
a stable fixed point is interpreted as an evidence of a
fluctuation-induced first-order phase transition in the system under
consideration. For models with no exact solutions or
rigorous proofs of the existence of a second order phase transition,
that is to say, for the majority of realistic models in
statistical physics, RG provides a tool to check the order of
the transition.

Now let us define the kind of problem we are going to discuss in this lecture.
It can be formulated as: {\it What is the order of
the normal-to-superconducting phase transition?}.
According to the
Bardeen-Cooper-Schriefer theory of the superconductivity, the
normal-to-superconducting  (NS) phase transition is a classical second
order phase transition described by the Landau-Ginsburg Hamiltonian
with a complex order parameter corresponding to the wave function
of the Cooper pairs. Taking into account the fluctuations of the order
parameter one can find values of corresponding critical exponents
which in this case will coincide with the critical exponents of $O(n)$
symmetric field-theoretical model for the case $n=2$ (the XY model).
Consequently, this leads to the assertion that the NS phase transition is
described by the same set of critical exponents as the phase
transition in the normal-to-superfluid liquid. The latter set of exponents
has been measured to a high accuracy\refnote{\cite{Lipa96}} and also calculated
by various methods\refnote{\cite{Baker78,LeGuillou80,Bagnuls85,Bagnuls87,%
Schloms87,Schloms89,Schloms90,Vakarchuk78}}.

Taking into account that the corresponding ``superfluid
liquid" is charged in case of the NS transition, however, complicates the
problem. This has been considered first of all by B.~I.~Halperin,
T.~C.~Lubensky and S.~Ma\refnote{\cite{Halperin74}} and since then different
ways of tackling it have been suggested. We shall discuss some of them
briefly in the subsequent section.

From an experimental viewpoint, when the question of the
order of the superconducting phase transition has been first discussed
it has been considered more or less academic since,
due to the large correlation length $\xi_0\sim 10^3$\AA, the
first order characteristics can be
seen only very near the phase transition; otherwise mean field behaviour is to
be expected. This situation has changed after the discovery of
high-T$_c$  superconductors with
correlation lengths of the order of lattice distances
($\xi_0\sim 1$ \AA)\refnote{\cite{Lobb87}}. Since then, critical effects have
been  observed
in several experiments
\refnote{\cite{Inderhees88,Salamon88,Inderhees91,Regan91,Mozurkewich92,
Salamon93,Junod93,Overend94}}.

Our main result presented in this lecture is that within the
framework of the RG method applied to the original superconductor model,
minimally coupled to the gauge field\refnote{\cite{Halperin74}},
one still can demonstrate the existence of a {\it
second order phase transition with critical exponents distinct
from those of a superfluid liquid}. To prove this we shall consider
the two-loop renormalization group functions
for the model, paying particular
attention to the fact that the loop expansion is asymptotic
\refnote{\cite{LeGuillou77,Lipatov77,Brezin78}}.
In this way we find several fixed points with new
scaling exponents and a rich crossover behavior. Some of our results
have been previously published in\refnote{\cite{Kolnberger90,Folk96,Folk97}}.

The lecture is, therefore, organized as follows. In the following section
we shall give a brief review of the methods used to study the problem we
are interested in. Then we describe the model of a
superconductor, provide results of its study by the mean field
approximation, obtain the expressions for the
renormalization group functions in a two loop approximation and
describe the results obtained subsequently without applying any
resummation procedure. Then in the next section we shall discuss
several well-established examples
in the modern theory of critical phenomena
where the resummation of the asymptotic series is applied.
After that we shall
present the pivotal section of our investigation: it is devoted to a study of
the RG functions and of the corresponding flows on the basis of the resummation
technique. We obtain as a result that a stable fixed
point is present, which is an evidence of a second order
phase transition in the model.  In the remainder of the
paper the asymptotic and effective values for the
critical exponents are calculated and we give expressions for the
amplitude ratios.  To conclude, these results are discussed in the closing
 section.

\section{2 NORMAL-TO-SUPERCONDUCTING TRANSITION:
1st OR 2nd ORDER?  \label{II}}

The order of a phase transition may have severe effects on
the physical quantities of a material. This is illustrated by the first-order
liquid-gas transition phenomena like overheating and undercooling
connected with the metastability at the transition. For second order phase
transitions, divergences in physical quantities occur (in the
thermodynamic limit, of course) leading to a dramatic increase of
the specific heat or the scattering of the light (the critical opalescence)
near the liquid-gas critical point.  Similar dramatic changes are associated
with the phase transitions which have occurred in the early stages of the
universe
where the questions discussed here for superconductors also find relevance
(for a recent review see\refnote{\cite{Meyer96}}).

As mentioned in the Introduction, the question of the
NS phase transition order becomes complicated when one accounts
for the fluctuations of the order parameter being coupled to the
"gauge field" (the vector potential of the fluctuating magnetic field
created by the Cooper pairs), the fluctuations of which also diverge at long
distances. Posed for the first time more than 20 years
ago\refnote{\cite{Halperin74}}, this problem has remained a challenging one in
the physics of the superconductivity until now.

The theoretical model of Halperin, Lubensky, and Ma
describing the relevant critical behaviour is the usual $O(n)$ symmetric
 $\phi^4$ model
with the $n/2$-component complex field $\phi$ coupled to a gauge field
which describes the fluctuating magnetic field created by Cooper pairs.
The answer obtained in\refnote{\cite{Halperin74}} states that due to the
coupling to the gauge field, in the mean field approximation a third order
term appears in the free energy of the superconductor and the NS phase
transition is of first order. This mean field analysis is
appropriate for the type-I superconductors\refnote{\cite{note1}} where the
fluctuations in the order parameter have no significant
effect on the thermodynamics of the transition.

The case of type-II superconductors is more
complicated, however, because fluctuations cannot be neglected.
Studying the problem within Wilson-Fisher recursion
relations\refnote{\cite{Wilson72}} in the first order of $\varepsilon$ it has
been
found\refnote{\cite{Halperin74}} that a stable fixed point (necessary, but not
sufficient for a second order phase transition) exists only for the
order parameter components number $n>365.9$, far exceeding
the superconductor case $n=2$.  The crossover near the first
order phase transition has been examined\refnote{\cite{Chen78}} and the
expression for the crossover function of the specific heat is given by the
one loop order perturbation theory.

The kinetics of fluctuations arising from vortex pairs in a superconductor
has been studied by means of numerical simulations\refnote{\cite{Filippov94}}.
The result leads to the conclusion that there exist nucleation processes
typical for the first order phase transition, confirming the mean field and RG
results of\refnote{\cite{Halperin74}}.
Note, though, that the mean field analysis applied to the Ginzburg-Landau
free energy of a superconductor\refnote{\cite{Malbouisson98}}
including a Chern-Simons term leads
to quantitatively different behaviour: for different values of the
topological mass, there occurs in a system either a fluctuation-induced first
order phase transition or only a second-order transition.
This result is also confirmed in the framework of one-loop RG
calculations\refnote{\cite{Malbouisson98}}.

The occurrence of a first order
phase transition has been also found in the massless scalar
electrodynamics\refnote{\cite{Coleman73,Kang74}} and confirmed to a linear
order in  $\varepsilon$ for the $n$-component Abelian Higgs models by
an explicit construction of the coexistence curve and the equation of
state\refnote{\cite{Lawrie82}}. In addition, the CP$^{N-1}$ non-linear sigma
model,
being related to the superconductor model in the limit of infinite charge
by means of the ($2+\varepsilon$) expansion, has been
shown\refnote{\cite{Hikami79}}
to posses a similar behaviour observed further by\refnote{\cite{Halperin74}}.

Results for type-II superconductors obtained by $\varepsilon$
-expansion methods\refnote{\cite{Halperin74,Chen78,Lawrie82}} appear to be
stable against the influence of different physical factors such as
the possibility of another (non-magnetic) ordering,
the presence of a disorder and a crystal anisotropy
when the study is performed by means of a strict
expansion. The scaling behaviour of a superconducting system
with an additional non-magnetic ordering studied by
$\varepsilon$-expansion methods provides one more example of a system where a
weak first-order phase transition occurs\refnote{\cite{Tonchev81}}.
The analysis of the influence of quenched impurities on the critical
behaviour of superconductors when taking account of the magnetic field
fluctuations demonstrates\refnote{\cite{Boyanovsky82,Uzunov83}}
the appearance of a
new stable fixed point for $1<n<366$. It has been
shown\refnote{\cite{Uzunov83}}, however, that this describes the critical
behaviour
in the range of space dimensionalities $d_c(n)<d<4$ with $d_c(2)=3.8$ and
results in a first order phase transition.

The RG flow for the superconductor model with quenched impurities
has been found\refnote{\cite{Athorne86}} to exhibit a stable focus surrounded
by an unstable limit cycle. The second order phase transition behaviour is
found to show up inside the limit cycle. Introducing random fields
with short and long range correlations does not lead to a second order
behaviour in the region of $(d,n)$ near $(3,2)$\refnote{\cite{Busiello86}}.
Note however, that studies of the influence of
quenched and annealed gauge fields on the spontaneous symmetry
breaking, performed in terms of Helmholtz free energy\refnote{\cite{Lovesey80}}
lead to the conclusion that in the first nontrivial or one-loop
approximation in the annealed model the spontaneous symmetry breaking
occurs through a first order transition for $d=2,3$ whereas the
quenched model displays a continuous phase transition. A more
complicated account of fluctuations in the annealed model changes the
nature of the transition to a continuous one, but the spontaneous
symmetry breaking is absent in the model with a quenched
disorder\refnote{\cite{Lovesey80}}.
The combined influence of the crystal anisotropy, the magnetic fluctuations,
and the quenched randomness on the critical behaviour of unconventional
superconductors\refnote{\cite{Blagoeva90}} studied by means of the RG analysis
within the $\varepsilon$-expansion\refnote{\cite{Busiello91}} gives
that only fluctuation-induced first order transitions
should occur in unconventional superconductors in the vicinity of the
critical point.

The results of mean field calculations, however, have been questioned by
Lovesey\refnote{\cite{Lovesey80}}, mentioned above, who has shown
that taking into account the gauge field fluctuations in the
calculation of the free energy leads back to a second order phase
transition. A further indication of a second order phase transition
has come several years later when this question has been studied as a lattice
problem by means of Monte-Carlo calculations and duality
arguments\refnote{\cite{Dasgupta13}}.
These results have confirmed that there are scenarios of the NS transition
that differ  from
those obtained in\refnote{\cite{Halperin74}}. Namely, the NS transition has
been found to be of second order, asymptotically equivalent to that of a
superfluid with a reversed temperature axis. Subsequent MC
simulations\refnote{\cite{Bartolomew83}} performed in different regions of
couplings lead to the result that the NS transition is strongly first
order deep in the type-I region and becomes more weakly first order
moving in the direction of the type-II region.  Beyond a certain point
the data reported in\refnote{\cite{Bartolomew83}} suggests a second-order
transition.  The
corresponding $O(n)$ nonlinear $\sigma$-model coupled to an Abelian
gauge field studied near two dimensions by $2+\varepsilon$
expansion\refnote{\cite{Lawrie83}} does not show a first order phase transition
either.

The existence of a tricritical point, where the order of the phase
transition changes from second into first, has been
predicted\refnote{\cite{Kleinert82}}
by representing the $3d$ superconductor model by a disordered field theory.
The position of the tricritical point is located slightly in the type-I
region for the values of the Ginzburg parameter\refnote{\cite{note1}}
$k<0.8/\sqrt{2}$. Starting from the dual formulation of the
Landau-Ginzburg theory, by means of the RG arguments, it has been shown that
the critical exponents of the NS transition  coincide with
those of a superfluid transition with a reversed temperature
axis\refnote{\cite{Kiometzis94}}.
But, while the correlation length critical exponent of the
normal-to-superconducting transition is predicted to coincide with
the ordinary 3D XY model, the divergency of the renormalized
penetration depth is characterised by the mean field value
$\nu=1/2$\refnote{\cite{Kiometzis94}}.

The influence of the critical fluctuations on the order of the NS transition
has been reconsidered on the basis of the field theoretical RG ideas
in\refnote{\cite{Kolnberger90}}. Here the two-loop flow
equations\refnote{\cite{Kolnberger90}} for the static parameters and the
$\zeta$-functions\refnote{\cite{Ford92}} are obtained and it is indicated that
a stable fixed point possibly exists and, as a consequence, a second
order phase transition may occur. An attractive feature of the flow
found in\refnote{\cite{Kolnberger90}} is that it discriminates between type-I
and type-II superconductors, depending on the initial (background)
values of the couplings. For small values of the ratio (coupling to
the gauge field)/(fourth order coupling) (appropriate for type-II
superconductors) the flow comes very near to the fixed point of the
uncharged model but ends in a new superconducting fixed point. For
large values of the ratio (type-I superconductors) the flow runs away.
For intermediate values of the ratio, the critical behavior may be
influenced by a second (unstable) superconducting fixed point with
scaling exponents quite different from those for the uncharged model.

A flow picture qualitatively similar to\refnote{\cite{Kolnberger90}} has been
obtained in\refnote{\cite{Bergerhoff96}} by solving the model of a charged
superconductor approximately with the help of nonperturbative flow equations,
a method which appears to give very encouraging results for critical
scalar field theories\refnote{\cite{Tetradis94,Graeter95}}. Depending on the
relative strength of the ratio (coupling to the gauge field)/(fourth
order coupling) a first or a second order phase
transition has been found.
An approximate description of the tricritical behaviour
has been given as well as an estimate of the correlation length critical
exponent $\nu$ and the pair correlation function critical exponent
$\eta$, which give us the second order phase transition has been
reported. Depending on three different assumptions for the stable
fixed point value of the coupling
to the gauge field, the following values are obtained
in two successive truncations of the potential;  $(\eta,\nu)$ = [(-0.13,
0.50); (-0.20, 0.47)], [(-0.13, 0.53); (-0.17, 0.58)], [(-0.13, 0.59);
(-0.15, 0.62)], indicating that the critical
exponents belong to the physical region $\eta>2-d$ and $\nu>0$,
independent of the truncation, clearly pointing towards a second order phase
 transition.

In the context of baryogenesis the question of the NS
phase transition order has been considered within the two loop approximation
in\refnote{\cite{Arnold94}} and the effective potential has been calculated.
The $\varepsilon$-expansion has been applied to the electroweak phase
transition in order to estimate various parameters of it in leading
and next-to-leading orders in $\varepsilon$, including the scalar
correlation length, the latent heat, the surface tension, the free energy
difference, the bubble nucleation rate, and
the baryon nonconservation rate. Of course, the transition has been found
to be a first order
since only run away flows occur in the strict
$\varepsilon$-expansion
perturbation theory.  Note, that in the electroweak scenario of
the baryogenesis there exists a so-called Sakharov requirement which is met
when the transition is strongly first order rather than second
order.

The NS transition problem has been also studied
by an analytical method which is not based on $\varepsilon$ or $1/n$
expansions. Using a non-perturbative method of solving
the approximate Dyson equation for arbitrary $d$ and $n$\refnote{\cite{Bray74}}
it has been found\refnote{\cite{Radzihovsky95}} that the NS phase transition is
governed by a "charged" fixed point. The value of the pair correlation function
critical exponent $\eta$ at $d=3,n=2$ is
$\eta(3,2)=-0.38$. It is interesting to note that although the result
for $\eta$ appears to be a well-behaved function of $d$ and $n$, it
breaks down at the critical value $n_c\simeq 18$ when expanded in
$\varepsilon$. Hence, the conclusion is drawn that the results of
the $\varepsilon$-expansion obtained in\refnote{\cite{Halperin74}} and, in
particular, the absence of a stable fixed point solution for $n<n_c\simeq
365.9$ are to be interpreted as the breakdown of the
$\varepsilon$-expansion rather than a fluctuation-induced
first-order phase transition. On the other hand, near $d=4$ the
results of\refnote{\cite{Radzihovsky95}} are in a good agreement with the
$\varepsilon$-expansion data\refnote{\cite{Halperin74}} for high $n$ ($n>366$).

Recently, the same problem studied with the RG technique in a fixed
dimension $d=3$ within the one-loop approximation has given an evidence of an
attractive charged fixed point distinct from that of a neutral
superfluid, leading, in particular, to the correlation length critical
exponents values $\nu \simeq 0.53$ and $\eta \simeq -0.70$
\refnote{\cite{Herbut96}}. Considered in the form of the continuum dual
theory\refnote{\cite{Herbut97}}, however, the magnetic penetration depth has
been shown to diverge with the XY exponent, contradicting results mentioned
above\refnote{\cite{Kolnberger90,Bergerhoff96,Radzihovsky95,Herbut96}}.
To investigate this discrepancy, MC simulations of the 3D isotropic
lattice superconductor in a zero external magnetic field have been performed.
This results in the conclusion that there is a single diverging length scale
consistent with the universality of the ordinary 3D XY
model\refnote{\cite{Olsson98}}.
Further applications of the model containing coupling to the  gauge
field have been suggested in the context of the quantum Hall
effect\refnote{\cite{hall}}.

Now, let us give a brief resume of the experimental data relevant to our
study. As mentioned in the Introduction, the effects of
thermodynamic fluctuations are generally small in conventional
low-$T_c$ superconductors because of their low transition temperatures
and large coherence length. In contrast, high transition
temperatures and small coherence lengths mean that critical fluctuations
are relevant in high-$T_c$ superconductors.  Though critical
fluctuations in high-$T_c$ superconductors have been observed in a series
of experiments ( see e.g.\refnote{\cite{Inderhees88,Salamon88,Inderhees91,%
Regan91,Mozurkewich92,Salamon93,Junod93,Overend94}}) their
interpretation has been changed somewhat. Deviations from the mean field
(i.e. first order) behaviour have accounted for either by $3d$
Gaussian fluctuations (giving, in particular, values for the
specific heat critical exponent $\alpha$ and the correlation length
critical exponent $\nu$:
$\alpha=\nu=1/2$)\refnote{\cite{Inderhees88,Inderhees91}}
or by a nontrivial XY behaviour
characteristic for an uncharged superfluid (with $\nu\simeq 2/3$ and
logarithmic divergences in
$\alpha$)\refnote{\cite{Salamon88,Regan91,Salamon93,Overend94}}.
Measurements of the heat
capacity\refnote{\cite{Salamon93,Overend94}}, the magnetization and the
electric conductivity\refnote{\cite{Salamon93}} of single-crystal samples of
$YBa_2Cu_3O_{7-x}$ in a magnetic fields near $T_c$  support the
existence of a critical regime governed by the XY-like critical
exponents\refnote{\cite{Salamon88,Regan91,Salamon93,Overend94,Lawrie94}}, a
similar conclusion follows from the crossover analysis of the
zero-field heat capacity on a comparable sample\refnote{\cite{Mozurkewich92}}.
The maximum applied magnetic field for which the 3D XY
scaling is valid, however, differs for various
materials\refnote{\cite{Nanda98}}.

To conclude this brief review it is worth mentioning one more physical
interpretation of a charged field  coupled to the gauge vector
potential. Namely, this is the nematic-smectic-A transition in liquid
crystals
\refnote{\cite{deGennes72,Halperin74a,Lubensky78,Kasting80,%
Anisimov90,Garland94}}.
The nematic phase is an orientationally ordered but
translationally disordered phase, rodlike molecules are aligned with
their long axes parallel to the director and the smectic-A phase
contains layers of molecules with their long axes perpendicular to the
layer. It has been proposed\refnote{\cite{deGennes72,Halperin74a}}
that this transition can be described by a  model similar to those
describing the NS transition in the charged case\refnote{\cite{Halperin74}}.
Now the smectic order parameter (being a complex field $\Psi(r)$ that
specifies the amplitude and the phase of the density modulation induced
by the layering)
is coupled to the director fluctuations.
Contrary to the NS transition, the nematic-smectic A transition
is characterized by a critical region in the experimentally
accessible range.
For certain materials it has  been indeed shown\refnote{\cite{Anisimov90}}
that both the latent heat data obtained through an adiabatic scanning
calorimetry as well as independent interface velocity measurements
near the Landau tricritical point can be fitted by a crossover function
consistent with a mean field free energy density that has a cubic
term\refnote{\cite{Halperin74}}, implying that the nematic-smectic-A transition
is a weakly first order. Many liquid crystals, though, appear
to exhibit a continuous nematic-smectic-A transition
(see\refnote{\cite{Garland94}} and references therein). High-resolution
heat-capacity and x-ray studies of the nematic-smectic-A transition
performed during the past twenty years (see\refnote{\cite{Garland94}} for a
comprehensive review) show complex systematic trends to crossover,
from three-dimensional XY to tricritical behaviour and an
anisotropic behaviour due to a coupling between the smectic
order parameter and director fluctuations.

\section{3 THE MODEL AND ITS "NAIVE" ANALYSIS \label{III}}

Now it is well-known, that the influence of the order parameter
fluctuations on the NS transition can be described by the
Landau-Ginsburg free energy functional:

\bigskip

\begin{equation}
F[\phi] = \int {\rm d}^{3}x \{ \frac{t_0}{2}|\phi_0|^{2}
+ \frac{1}{2}|(\nabla \phi_0|^2
+ \frac{u_0}{4!}|\phi_0|^4 \} ,
\label{1}
\end{equation}

\bigskip

\noindent
$t_0$ is temperature-dependent, $u_0$ is a coupling constant and
the complex order parameter $\phi_0$ is connected with the wave function
of Cooper pairs. The Cooper pairs are charged and therefore create a fluctuating
magnetic field which leads to the appearance of additional terms in
the free energy functional. Note, that this is not the case of a
normal-to-superfluid transition in a neutral (uncharged) fluid, which is
well described by (\ref{1}) without any modification. Describing the
fluctuating magnetic field $\bf B$ by the vector potential {\bf A}
(${\bf B} = {\rm rot} {\bf A}$) and adding to (\ref{1}) the minimal
coupling between the fluctuating vector potential and the order parameter
one obtains the free energy functional $F[\Psi, {\bf A}]$,
originally considered in\refnote{\cite{Halperin74}} for a generalized
superconductor in $d$ dimensions with a $d$-dimensional vector
potential ${\bf A}$ and the order parameter $\Psi$ consisting of $n/2$
complex components.

One can now describe the fluctuation effects by an Abelian Higgs
model with the gauge invariant Hamiltonian\refnote{\cite{Halperin74}}:

\bigskip

\begin{equation}
H = \int {\rm d}^{d}x \{ \frac{t_0}{2}|\Psi_0|^{2}
+ \frac{1}{2}|(\nabla - ie_0{\bf A}_0)\Psi_0|^2
+\frac{u_0}{4!}|\Psi_0|^4 +
\frac{1}{2}(\nabla \times {\bf A}_0)^2 \},
\label{2}
\end{equation}

\smallskip

\noindent
which depends on the bare parameters $t_{0}$, $e_{0}$, $u_{0}$. The parameter
$t_{0}$ changes its sign at some temperature, the rest of the parameters
are considered  temperature-independent. When the coupling constant
$e_0=0$ no magnetic fluctuations are induced and the model reduces to
the usual field theory (\ref{1}) describing a second-order phase
transition and corresponding in the particular case $n=2$ to the
superfluid transition in $^4$He.

The mean field results for the critical behaviour of the model with a
free energy functional $F[\Psi, {\bf A}]$ corresponding to the
Hamiltonian (\ref{2})
have been already reported in the
original paper of Halperin, Lubensky and Ma\refnote{\cite{Halperin74}}. In the
framework of the mean field theory one can determine that systems
characterized by the free energy functionals $F[\phi]$  (\ref{1}) and
$F[\Psi, {\bf A}]$ ($n=2$) possess a qualitatively different critical
behaviour.  Neglecting the order parameter fluctuations (in
accordance with the Ginzburg criterion this may be done for a good
type-I superconductor) shows that depending on the sign of $t_0$
the free energy (\ref{1}) is minimized by the value of
the order parameter $\phi=0$ for $t_0 > 0$
or by a non-zero value, when $t_0 < 0$,
and that the appearance of the
non-zero order parameter is continious; in the system under
consideration a second order phase transition occurs.

When applied to the free energy functional $F[\Psi,{\bf A}]$, however,
the mean field theory predicts a qualitatively different behaviour.
Defining the effective
free energy $F_{\rm eff}[\Psi]$  as a function of the single variable
$\Psi$ by taking the trace over the configurations of the vector
potential ${\bf A}$
one finds\refnote{\cite{Halperin74}} that the expression for
$F[\Psi]$ will contain a term which has a negative sign and is
proportional to $|\Psi|^3$. Such a term inevitably leads to a first
order transition; $F_{\rm eff}[\Psi]$ develops a minimum at a finite
value of
$\Psi$ when the coefficient of the quadratic term is still slightly
positive.

As it has been already mentioned, the above reasoning is appropriate
for a type-I superconductor. The case of type-II superconductors is
considerably more complicated. Here, fluctuations in $\Psi$ cannot
be neglected and one must choose an appropriate technique to study the
problem. Originally, the critical behaviour of the model (\ref{2}) in the
presence of order parameter fluctuations has been studied
in\refnote{\cite{Halperin74}} with the help of Wilson-Fisher recursion
relations\refnote{\cite{Wilson72}} in the first order of $\varepsilon=4-d$
and in result it is shown, in particular, that the second order phase
transition is absent for $n=2$ in the region of couplings appropriate for the
type-II superconductor.  We will reproduce these $\varepsilon$-expansion results
below and then continue to analyze the problem further.

In order to describe the long-distance properties of the model (\ref{2})
arising in the vicinity of the phase transition point we shall use
a field-theoretical RG approach. Two-loop
results\refnote{\cite{Kolnberger90}} for the RG functions
corresponding to (\ref{2}) are obtained on the basis of a dimensional
regularization and a minimal subtraction scheme\refnote{\cite{tHooft72}},
defining the renormalized quantities so as to subtract all poles at
$\varepsilon=4-d=0$ from the renormalized vertex functions.
The renormalized fields, mass and couplings are introduced by:
\begin{eqnarray}
&&\Psi_0 = Z_{\Psi}^{1/2}\Psi, \qquad
{\bf A}_0 = Z_{\rm A}^{1/2}{\bf A},  \qquad
t_0 - t_{0c}= Z_t Z_{\Psi}^{-1}t,
\nonumber \\
&&e_0^2 = Z_e^{1} Z_{\rm A}^{-1}Z_{\Psi}^{-1} e^2 
\mu^{\varepsilon}S_d^{-1}, \qquad
u_0=Z_u Z_{\Psi}^{-2} u \mu^{\varepsilon}S_d^{-1}
\label{3}
\end{eqnarray}
with $\varepsilon=4-d$. Here, $\mu$ is an external momentum scale,
$t_{0c}$
is a shift which for the results considered here can be set to
zero, and $S_d$ stands for the surface of a $d$-dimensional hypersphere:
$S_d=2^{1-d}\pi^{-d/2}/\Gamma(d/2)$. The $Z$-factors are determined by
the condition that all poles at $\varepsilon=0$ are removed from the
renormalized vertex functions.

The RG equations are written bearing in mind the fact that the bare vertex
functions $\Gamma_0^{N,M}$ are calculated with the help of the bare Hamiltonian
(\ref{2}) as a sum of one-particle irreducible (1PI)
diagrams\refnote{\cite{note1a}}:

\bigskip

\begin{equation} \label{3a}
\Gamma_0^{N,M} (\{ r \}, \, \{ R \}) =
<\Psi_0(r_1)\dots \Psi_0(r_N)A_0(R_1)\dots A_0(R_M)>_{\rm 1PI},
\end{equation}

\smallskip

\noindent
which do not depend on the scale $\mu$ and their derivatives with respect to
$\mu$ at fixed bare parameters are equal to zero. So one gets

\bigskip

\begin{equation} \label{3b}
\mu \frac{\partial}{\partial \mu} \Gamma_0^{N,M}|_0=
\mu \frac{\partial}{\partial \mu} Z_{\Psi}^{-N/2}
Z_{A}^{-M/2}\Gamma_R^{N,M}|_0=0,
\end{equation}

\smallskip

\noindent
where the index $0$ means a differentiation at fixed bare parameters.
So, the RG equations for the renormalized vertex function
$\Gamma_R^{N,M}$ will be:

\bigskip

\begin{equation} \label{3c}
\Big ( \mu \frac{\partial}{\partial \mu} +
\beta_u \frac{\partial}{\partial u} +
\beta_f \frac{\partial}{\partial f} +
\zeta_{\nu} t  \frac{\partial}{\partial t} -
\frac{N}{2} \zeta_{\Psi} -
\frac{M}{2} \zeta_{A} \Big )
\Gamma_R^{N,M}(t,u,f,\mu) =0,
\end{equation}

\smallskip

\noindent
where $f=e^2$, $\zeta_{\nu} = \zeta_{\Psi} - \zeta_t$ and the RG
functions read

\bigskip

\begin{eqnarray} \nonumber
\beta_u(u,f) = \mu \frac{\partial u}{\partial \mu}  |_0, \qquad
\beta_f(u,f) = \mu \frac{\partial f}{\partial \mu}  |_0, \\
\zeta_{\Psi} = \mu \frac{\partial \ln Z_{\Psi}}{\partial \mu} |_0,
\qquad
\zeta_{A} = \mu \frac{\partial \ln Z_{A}}{\partial \mu} |_0,
\qquad
\zeta_{t} = \mu \frac{\partial \ln Z_{t}}{\partial \mu} |_0.
\label{3d}
\end{eqnarray}

\smallskip

\noindent
Using the method of characteristics the solution of the RG
equation may be written formally as:

\bigskip

\begin{equation}
\Gamma_R^{N,M}(t,u,f,\mu) = X(l)^{N/2} (X^{\prime}(l))^{M/2}
\Gamma_R^{N,M}(Y(l)t,u(l),f(l),\mu l),
\label{3e}
\end{equation}
\smallskip

\noindent
where the characteristics are the solutions of the ordinary
differential equations:

\bigskip

\begin{eqnarray} \nonumber
l \frac{d}{d l} \ln X(l) = \zeta _{\Psi} (u(l), f(l)), \qquad
l \frac{d}{d l} \ln X^{\prime}(l) = \zeta _{A} (u(l), f(l)), \\
\label{3f}
l \frac{d}{d l} \ln Y(l) = \zeta _{\nu} (u(l), f(l)),
\;\; \qquad \qquad \qquad \qquad \qquad \qquad \qquad \\
l \frac{d}{d l} u(l) = \beta _{u} (u(l), f(l)), \qquad
l \frac{d}{d l} f(l) = \beta _{f} (u(l), f(l)) \qquad \qquad
\label{9}
\end{eqnarray}
\smallskip

\noindent
with
\begin{equation}
X(1)=X^{\prime}(1)=Y(1)=1, \qquad u(1)=u, \qquad f(1) = f.
\end{equation}
For small values of $l$, the equation (\ref{3e}) is mapping the large
length scales (the critical region) to the noncritical point $l=1$.
In this limit the scale-dependent values of the couplings $u(l),
\, f(l)$ will approach the stable fixed point, if it exists.

The fixed points $u^*, \, f^*$ of the differential equations
(\ref{9}) are given by the solutions of the system of equations:
\begin{eqnarray} \nonumber
\beta_f(u^*,\, f^*) &=& 0,
\\
\beta_u(u^*, \, f^*) &=&  0.
\label{9b}
\end{eqnarray}
The stable fixed point is defined as the fixed point where the
stability matrix
\begin{equation}
B_{ij}= \frac{\partial \beta_{u_i}}{\partial u_j} \, ,
\qquad u_i=\{ u, \, f\}
\label{stab}
\end{equation}
possess positive eigenvalues (or if complex, the eigenvalues with positive real
parts). The stable fixed point corresponds to the critical
point of the system. As we have mentioned above,
in the limit $l\rightarrow0$ (corresponding to
the limit of an infinite correlation length) the renormalized couplings
reach the values they have in the stable fixed point.

Now we can write the results for the RG functions obtained in a
two-loop ap\-pro\-xi\-ma\-ti\-on\refnote{\cite{Kolnberger90}} following the
above described procedure in frames of the dimensional regularization and
the minimal subtractions schemes.
From a Ward identity one has
$Z_{\Psi}=Z_e$, and the remaining $Z$-factors are to be found from the
corresponding vertex functions $\Gamma^{2,0}$, $\Gamma^{0,2}$, and
$\Gamma^{4,0}$.  Since the gauge field is
massless, the renormalization has been performed at a finite wave
vector. The results in the two-loop order read:
\begin{eqnarray}
&&Z_{\Psi} = 1 + \frac{1}{\varepsilon}\{3e^2 - u^2(n+2)/144 +
e^4[(n+18)/4\varepsilon-(11n+18)/48]\},
\label{4}
\\
&&Z_{\rm A} = 1 + \frac{1}{\varepsilon}\{-ne^2/6 - ne^4/2\},
\label{5}
\\
&&Z_{t} =1 + \frac{1}{\varepsilon}\{
(n+2)u/6 + u^2[(n+2)(n+5)/36\varepsilon-(n+2)/24] +
\nonumber \\
&& ue^2 [-(n+2)(1/2\varepsilon -1/3)]+
e^4[(3n+6)/2\varepsilon+(5n+1)/4]\},
\label{6}
\\
&& Z_{u} = 1 + \frac{1}{\varepsilon}\{
(n+8)u/6 + 18e^4/u + u^2[(n+8)^2/36\varepsilon-(5n+22)/36]
+  \nonumber \\
&& ue^2[-(n+8)/2\varepsilon+(n+5)/3] + e^4[(3n+24)/\varepsilon
+(5n+13)/2] +
\nonumber \\
&& e^6/u[3(n+18)/\varepsilon -7n/2-45]\}.
\label{7}
\end{eqnarray}

Following the standard procedure one then shows that the expressions for
$\beta$-functions in the two-loop approximation will be:

\smallskip

\begin{eqnarray}
&&\beta_f = -\varepsilon f + \frac{n}{6} f^2 + n f^3,
\label{10}
\\
&&\beta_u = -\varepsilon u + \frac{n+8}{6} u^2 -
     \frac{3n+14}{12} u^3 - 6uf + 18 f^2
\nonumber \\
&&+\frac{2 n + 10}{3} u^2 f +
    \frac{71n+174}{12} u f^2 - (7n+90)f^3.
\label{11}
\end{eqnarray}

\smallskip

\noindent
The previous analysis of the equations of type (\ref{10}),
(\ref{11}) either on a one-loop\refnote{\cite{Halperin74}} or two-loop
level\refnote{\cite{Kolnberger90}} has been based on the direct solutions of
the equation for the fixed point. In the present study we want to point
out that the series have a zero radius of convergence
and they are known to be asymptotic at best. Therefore, some additional
mathematical methods have to be applied in order to obtain a reliable
information on their basis.

We shall start by recalling the results of an $\varepsilon^{2}$-expansion for
$\beta$-functions\refnote{\cite{Halperin74,Kolnberger90}}. For the second order
in $\varepsilon$ one obtains three fixed points: Gaussian
($u^{*G}=f^{*G}=0$),``Un\-char\-ged" ($u^{*U}\neq 0, f^{*U}=0$) and
``Charged" ($u^{*C}\neq 0, f^{*C}\neq 0$), to be denoted as G, U,
C.  The expressions for them read:

\medskip

\begin{eqnarray}
\makebox{G}: & u^{*G}=0, & f^{*G}=0,
\label{12} \\
\makebox{U}: &
u^{*U}=u_{1}^{U} \varepsilon + u_{2}^{U} \varepsilon^2, & f^{*U}= 0,
\label{13} \\
\makebox{C}: & u^{*C}=u_{1}^{C} \varepsilon +
u_{2}^{C} \varepsilon^2, & f^{*C}=f_{1}^{C} \varepsilon + f_{2}^{C}
\varepsilon^2,
\label{14}
\end{eqnarray}

\smallskip

\noindent
where

\medskip

\begin{eqnarray*}
&&u_{1}^{U} = \frac{6}{n+8}, \qquad u_{2}^{U} = \frac{18
      (3n+14)}{(n+8)^3}\, ,
\\ &&
u_{1}^{C} = \frac{3(n+36) + (n^2 - 360 n - 2160)^{1/2}}{3n(n +8)}\, ,
\\
&& u_{2}^{C} = \frac{a_2}{a_1}, \qquad f_{1}^{C} = \frac{6}{n},  \qquad
f_{2}^{C} = - \Bigg(\frac{6}{n}\Bigg)^3 n\, ,
\end{eqnarray*}
with
\begin{eqnarray*}
&&a_1 = 1 + \frac{n + 8}{3} u_{1}^{C} - \frac{36}{n},
\qquad
a_2 = \frac{3n+14}{12}\Bigg(u_{1}^{C}\Bigg)^3
-6n u_{1}^{C}\Bigg(\frac{6}{n}\Bigg)^3 +
\\ &&
 36n \Bigg(\frac{6}{n}\Bigg)^4 -
     \frac{(n+5)4}{n} \Bigg(u_{1}^{C}\Bigg)^3 -
     \frac{3(71 n + 174)}{n^2} u_{1}^{C}
     +\Bigg(\frac{6}{n}\Bigg)^3 (7n+90).
\end{eqnarray*}

\medskip

Almost all physical results concerning the phase transition described by
the field theory (\ref{2}) are to some extent based on the
information given by (\ref{12}) - (\ref{14}).
The main ones read:

\begin{figure}[t]
\begin{centering}
\setlength{\unitlength}{1mm}
\begin{picture}(126,110)
\epsfxsize=126mm
\epsfysize=90mm
\put(0,0){\epsffile[22 20 822 582]{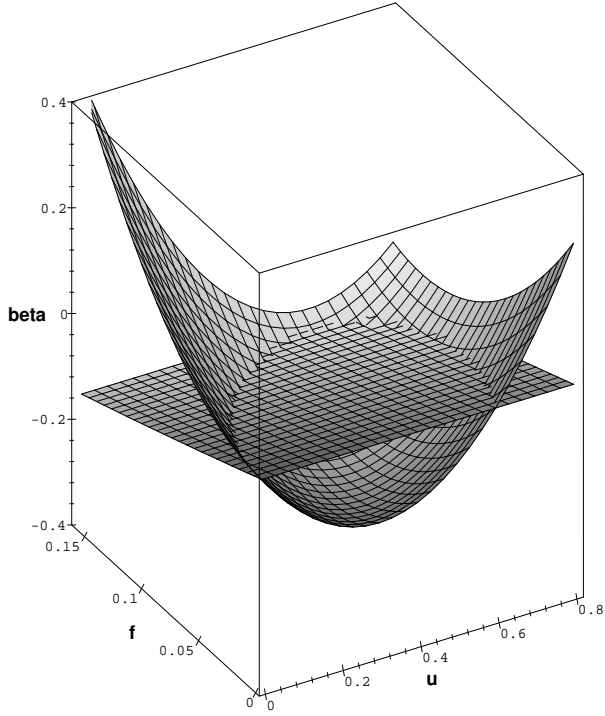}}
\end{picture}\\
\end{centering}
\caption{\label{fig1}
$\beta$-functions of the model of superconductor $\beta_u(u,f)$,
$\beta_f(u,f)$ in one-loop approximation for $d=3$, $n=2$. The fixed
points have coordinates ($u^*=f^*=0$), ($u^*=0.6, \, f^*=0$).
}
\end{figure}

\begin{itemize}
\item[(i)] the fixed point U is unstable with respect to the presence
of an $f$-symmetry at $d<4$ with a stability exponent
$$
\lambda_f (u=u^{*U}, f=f^{*U}=0) = \frac{\partial \beta_f}
{\partial f}|_{U} = -\varepsilon ;
$$
\item[(ii)] the fixed point C appears to be complex for $n<n_c=365.9$
already\refnote{\cite{Halperin74}} on a one-loop level. The 
stability exponent is given by
$$
\lambda_u (u=u^{*C}, f=f^{*C}) = \frac{\partial \beta_u}
{\partial u}|_{C}
$$
and on the two-loop level it reads:

\vspace{0mm}

$$
\lambda_u = - \varepsilon s,
\quad s=\Bigg[ \Bigg(1 + \frac{36}{n}\Bigg)^2-
\frac{432(n+8)}{n^2}\Bigg]^{1/2}
$$

\vspace{0mm}

\noindent
leading to an oscillatory flow of $u$ in a one-loop order
below $n_c$ with the so\-lu\-ti\-on\refnote{\cite{Chen78,Kolnberger90}}:
\begin{eqnarray}
&&f(l) = \frac{6 fl^{-\varepsilon}}
{6 + n\varepsilon f(l^{-\varepsilon}-1)},
\label{15}
\\
&&u(l)=f(l) \frac{n}{2(n+8)} \Bigg\{ s \tan \Bigg[\frac{s}{2}
\ln \Bigg(f(l)f^{-1}l^{\varepsilon}\Bigg)
\nonumber \\
&& + \arctan \Bigg(\frac{2(n+8)}{sn}
\frac{u}{f} + \frac {n+36}{ns}\Bigg)\Bigg]
- \frac {n+36}{n} \Bigg\};
\label{16}
\end{eqnarray}
here $f$ and $u$ are the initial parameters for $l=1$;
\item[(iii)] from the condition of the positiveness of the fixed point
coordinate $f^*$ ($f=e^2$) it follows that at $\varepsilon=1$,
$n$ has to be larger that $36$. This questions the applicability of
the $\varepsilon$-expansion for $n=2$ to $d=3$.
\end{itemize}

\begin{figure}
\begin{centering}
\setlength{\unitlength}{1mm}
\begin{picture}(126,110)
\epsfxsize=126mm
\epsfysize=90mm
\put(0,0){\epsffile[22 20 822 582]{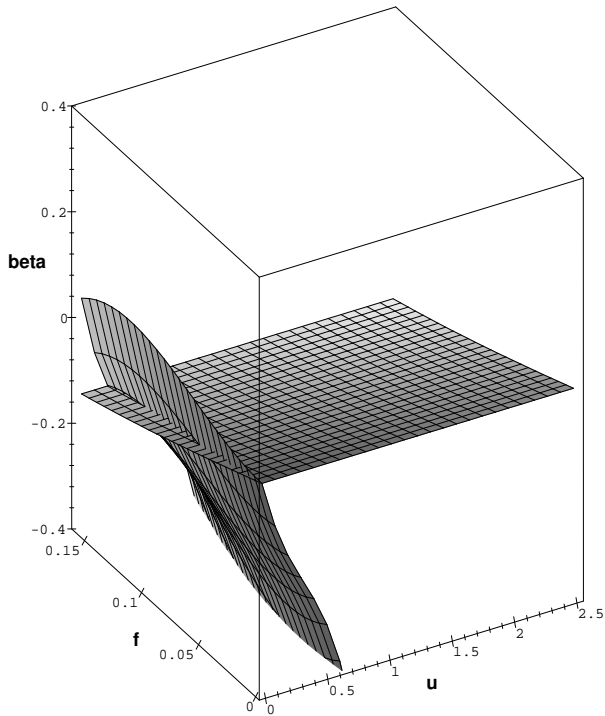}}
\end{picture}\\
\end{centering}
\caption{\label{fig2}
$\beta$-functions of the model of superconductor $\beta_u(u,f)$,
$\beta_f(u,f)$ in two-loop approximation for $d=3$, $n=2$.
Only the Gaussian fixed point $u^*=f^*=0$ survives.
}
\end{figure}

So, it follows that for the ``superconductor"
case $n=2$ which is the most interesting from a physical point of view
a stable fixed point does not exist and, therefore, the observed phase
transition is of a first order.

Now we will study the RG equations in the minimal
subtraction scheme in the framework of $d=3$
theory\refnote{\cite{Schloms87,Schloms89,Schloms90}} putting $\varepsilon=1$ in
expressions for the RG functions and studying the perturbation
theory in powers of coupling constants. The latter corresponds to the
number of loops in Feynman diagrams and one develops, therefore, the
perturbation theory in a successive number of loops. Direct calculations
based on equations (\ref{10}), (\ref{11}) at fixed $d=3$
do not bring
qualitatively new features to the described above analysis. In the
one-loop approximation, leaving square terms in (\ref{10}), (\ref{11})
one finds that only one nontrivial fixed point
$u^*= 6/(n+8), \, f^*=0$ exists. The $\beta$-functions $\beta_u(u,f)$,
$\beta_f(u,f)$ in the one-loop approximation at $d=3$, $n=2$
are shown in the Fig. \ref{fig1}.
The simultaneous intersection of the surfaces corresponding to both
functions with the plane $\beta=0$ results in the fixed points;
for the $n=2$ they have coordinates $u^*=f^*=0$ and $u^*=0.6,\,f^*=0$
as seen in the picture. In the two-loop approximation only
the Gaussian fixed point survives, as one may see from Fig.
\ref{fig2}.

Nevertheless one should note that such a straightforward
interpretation of the above expansions data has been questioned and the way
of analyzing the series for $\beta$-functions (\ref{10}), (\ref{11})
avoiding the strict $\varepsilon$-expansion and
exploiting the information on the accurate solution for the pure model
at $d=3$, has been proposed\refnote{\cite{Kolnberger90}}. Also, from the
comparison of $\varepsilon$-expansion data for $f^*$ (giving positive values of
$f^*$ only for $n>36$) with the value of $f^*$ obtained without the
$\varepsilon$-expansion (remaining positive for all $n$) it
has been conjectured that the lower boundary for $n$ resulting in
the negative $f^*$ may be an artifact of the expansion procedure.
Let us now consider expressions for the RG functions more carefully,
paying attention to their possible asymptotic nature and treating them
with a resummation procedure.

\section{4 RESUMMATION \label{IV}}

The resummation technique appropriate
for critical phenomena and applied to the asymptotic series for
the RG functions enables one to obtain extremely
accurate values of the critical exponents\refnote{\cite{note2}}.
In fact, the asymptotic nature of the series for the
RG functions has only been proven in the case of
the $\phi^4$ model containing one coupling of an $O(n)$-symmetry
(the $n$-vector model).  The high-order asymptotics for these series
are known\refnote{\cite{LeGuillou77,Lipatov77,Brezin78}} in analytical
form as well. These results give the possibility of obtaining precise
values for the critical exponents of the $n$-vector model by a resummation
of the corresponding series for the renormalization group functions
(see e.g.\refnote{\cite{Baker78,LeGuillou80,Schloms87}}).
As far as we know, no
information similar to that obtained
in\refnote{\cite{LeGuillou77,Lipatov77,Brezin78}}  for the
``uncharged" case ($f=0$) is available for the ``charged" model we are
 considering here.
For models containing several couplings of different symmetries the
asymptotic nature of the corresponding series for the RG
functions is a generally accepted belief rather than a proven fact.

As it will be important in the course of our future analysis
we  shall mention here the weakly diluted
$n$-vector model, describing the ferromagnetic ordering in a
system of $N_1$ classical $n$-component ``spins" located in $N$ sites
of a lattice ($N_1/N < 1$) and quenched in a certain configuration.
Using the replica trick in order to perform the quenched averaging
one concludes\refnote{\cite{Grinstein76}} that an effective Hamiltonian of such
a model contains two fourth order terms of a different symmetry and reads:

\bigskip

\begin{equation}
H = \int {\rm d}^{d}x
\Big\{
{1\over 2} \sum_{\alpha=1}^{m} \left[|\nabla \vec{\phi}^\alpha|^2+
m_0^2 |\vec{\phi}^\alpha|^2\right]- {v_{0}\over 8}
\left(\sum_{\alpha=1}^{m}|\vec{\phi}^\alpha|^2 \right)^2 +
{u_{0}\over 4!}
\sum_{\alpha=1}^{m}\left(|\vec{\phi}^\alpha|^2 \right)^2 \Big\},
\label{16a}
\end{equation}

\smallskip

\noindent
where $\vec{\phi}^{\alpha}$ is an $n$-component vector
$\vec{\phi}^\alpha=
(\phi^{\alpha,1},\phi^{\alpha,2},\dots,\phi^{\alpha,n})$;
$u_{0} > 0, v_{0} > 0$ are bare coupling constants; $m_0$ is the bare
mass and in the final results a replica limit $m \rightarrow 0$ is to
be taken. The RG functions for this model are obtained in the
form of a double series in renormalized couplings $u,\, v$ and
the asymptotic nature of the series has not been proven for
this model up till now\refnote{\cite{note3}}.
Nevertheless, the appropriate resummation technique (applied {\it as if}
these series are asymptotic ones) enables one to obtain accurate
values for critical exponents in three
dimensions\refnote{\cite{Jug83,Mayer89,Mayer89a,Holovatch92,Janssen95}}
and to describe
(in the $n=1$ case) the experimentally observed crossover to a new type of a
critical behavior caused by the weak
dilution\refnote{\cite{Mitchell86,Birgeneau88}}.
These results have been confirmed by
Monte-Carlo\refnote{\cite{Wang90,Wang94}} and Monte-Carlo
RG\refnote{\cite{Holey90}} calculations.

The two main ways of a resummation commonly used for the asymptotic series
arising in the RG approach are:
$(\bf {i})$ a resummation based on the conformal mapping technique and
$(\bf {ii})$ the Pad\'e-Borel resummation.
Case $(\bf {i})$ is based on the conformal transformation, which
maps part of the analytical domain containing the real positive
axis onto a circle centered at the origin and the asymptotic expansion
for a certain function is thus re-written in the form of a new series
(see\refnote{\cite{LeGuillou80}}). This resummation, however, is based on the
knowledge of subtle details of asymptotics (the location of the pole,
the high-order behavior) which are not available in our case.

In the absence of knowledge about the singularities of the series,
the most appropriate method which can be used to perform the
analytical continuation is the Pad\'e approximation resulting in the
Pad\'e-Borel resummation technique $(\bf {ii})$ (see
e.g.\refnote{\cite{Baker78}}).
We are going to apply it to the special case of $f=0$, so let us disucss it in
detail.

Starting from the Taylor series for the function $f(u)$:

\vspace{2mm}

\begin{equation}
f(u) = \sum_{j\geq0} c_j \ u^j,
\label{17}
\end{equation}

\vspace{2mm}

\noindent
one constructs the Borel-Leroy transform
\begin{equation}
F(ut) = \sum_{j\geq0} \frac {c_j}{\Gamma(j+p+1)} \ (ut)^j,
\label{18}
\end{equation}
with $\Gamma(x)$  the Euler's gamma-function and $p$ some
positive number\refnote{\cite{note4}}.
Then one represents (\ref{18}) in the form of the Pad\'e
approximant $F_{[L/M]}^{\rm Pad\acute{e}}(ut)$:
\begin{equation}
F_{[L/M]}^{\rm Pad\acute{e}}(x) = \frac
{\sum\limits_{i=0}^{L}a_ix^i}{\sum\limits_{j=0}^{M}b_jx^j}
\label{18a}
\end{equation}
(in the subsequent analysis, proceeding in
a two-loop approximation we will use the [1/1] Pad\'e approximant) and
the resummed function will be given by:
\begin{equation}
f^{\rm Res}(u) = \int\limits_{0}^{\infty} dt \  e^{-t} \ t^p
\ F^{\rm Pad\acute{e}}(ut) .
\label{19}
\end{equation}

The resummation scheme (\ref{18}) --  (\ref{19}) of the
(asymptotic) series in one variable (\ref{17}) is easily generalized to
the two-variable case when the series is given in a form:
\begin{equation}
f(u,v) = \sum_{j,j\geq0} c_{i,j} \ u^i \ v^j,
\label{19a}
\end{equation}

\vspace{2mm}

\noindent
with the Borel-Leroy transform
\begin{equation}
F(u,v,t) = \sum_{i,j\geq0} \frac {c_{i,j}}{\Gamma(i+j+p+1)} \
(ut)^i (vt)^j.
\label{19b}
\end{equation}
The procedure is aimed to help how to choose an appropriate form of the
analytic continuation of the series (\ref{19b}). Two most common
methods to proceed are the Borel resummation combined with Chisholm
approximants and the Borel resummation of the resolvent series,
presented in a form of the Pad\'e approximant. For the first method, in order
to write an analytic continuation of the series (\ref{19b}) one uses
the rational approximants of two variables:  so-called Canterbury
approximants or a generalized Chisholm
approximants\refnote{\cite{Chisholm73,Baker81}}
which are a generalization of Pad\'e
approximants in the case of several variables, representing
(\ref{19b}) in a form:
\begin{equation}
F^{\rm Chisholm}(u,v,t) = \frac
{\sum\limits_{i,j}a_{i,j}u^iv^jt^{i+j}}{\sum\limits_{i,j}b_{i,j}u^iv^jt^{i+j}},
\label{19c}
\end{equation}
(sums in the numerator and denominator are limited by the condition
of the correspondence between known numbers of terms in the initial series
and that in the approximant). Again, the resummed function is given by
an integral (\ref{19}):
\begin{equation}
f^{\rm Res}(u,v) =
\int\limits_{0}^{\infty} dt \  e^{-t} \ t^p \ F^{\rm Chisholm}(ut) .
\label{19d}
\end{equation}
Proceeding with the second method, one writes for the series of two variables
(\ref{19a}) the so-called resolvent series
${\cal F}(u,v,\tau)$\refnote{\cite{Watson74,Baker81}}
introducing an auxiliary variable $\tau$, which allows the separation of
contributions from different orders of the perturbation theory in the
variables $u, \, v$:
\begin{eqnarray}  \label{19e}
{\cal F}(u,v,\tau) &=& \sum_{i,j\geq0} c_{i,j} \,
(u\tau)^i \, (v\tau)^j,
\\ \nonumber
f(u,v) & = & {\cal F}(u,v,\tau=1).
\end{eqnarray}
Now the resummation of the series ${\cal F}(u,v,\tau)$ is performed
with respect to the variable $\tau$ as for the series in a single variable,
applying the above described scheme (\ref{18}) --  (\ref{19}).

\begin{figure}
\begin{centering}
\setlength{\unitlength}{1mm}
\begin{picture}(126,110)
\epsfxsize=126mm
\epsfysize=90mm
\put(0,0){\epsffile[22 20 822 582]{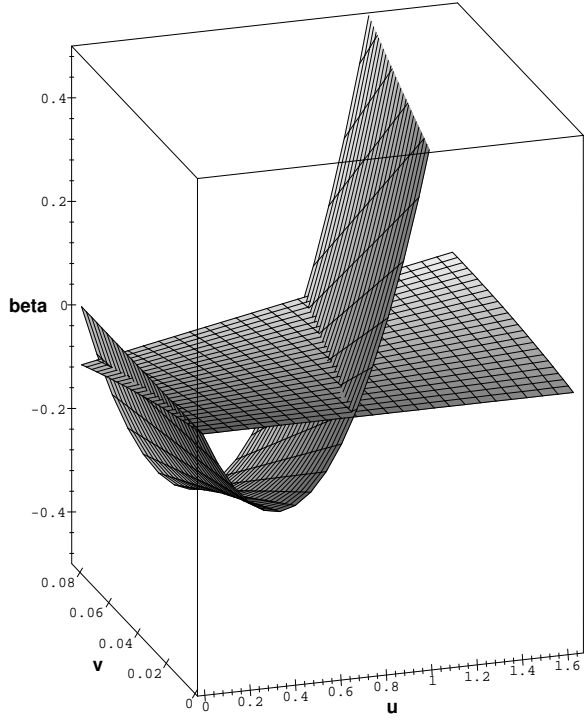}}
\end{picture}\\
\end{centering}
\caption{\label{fig3}
$\beta$-functions of the diluted Ising model
$\beta_u(u,v)$,
$\beta_v(u,v)$ in one-loop approximation for $d=3$. The fixed
points have coordinates ($u^*=v^*=0$), ($u^*=0.667, \, v^*=0$).
}
\end{figure}

Let us illustrate how the resummation procedure works in the case of
the effective Hamiltonian (\ref{16a}). In order to make a direct
comparison with the superconductor case, let us take the
$\beta$-functions obtained for the model (\ref{16a}) in the minimal
subtraction scheme in the two-loop approximation, though
high-order results are available for this
model\refnote{\cite{Janssen95,Kleinert95}} as well as
results\refnote{\cite{Jug83,Mayer89,Mayer89a,Holovatch92}} obtained in the
$d=3$ massive field theoretical approach\refnote{\cite{Parisi}}. The
expressions
for the $\beta$-functions, corresponding to the renormalized couplings $u,
\, v$ in the replica limit $m \rightarrow 0$ for the Ising model
(n=1) read:

\smallskip

\begin{eqnarray}
\beta_u &=& -\varepsilon u + \frac{3}{4} u^2 - 6uv -
\frac{17}{12} u^3 + \frac{23}{2} u^2v - \frac{41}{2} uv^2,
\label{19f}
\\
\beta_v &=& -\varepsilon v + uv+ -4 v^2 - \frac{5}{12} u^2 v +
     \frac{11}{2} uv^2 - \frac{21}{2} v^3.
\label{19g}
\end{eqnarray}

\smallskip

\noindent
We shall not present the expressions for the other RG functions here, as
we are going to study only the fixed point equations.

\begin{figure}[ht]
\begin{centering}
\setlength{\unitlength}{1mm}
\begin{picture}(126,110)
\epsfxsize=126mm
\epsfysize=90mm
\put(10,-10){\epsffile[22 20 822 582]{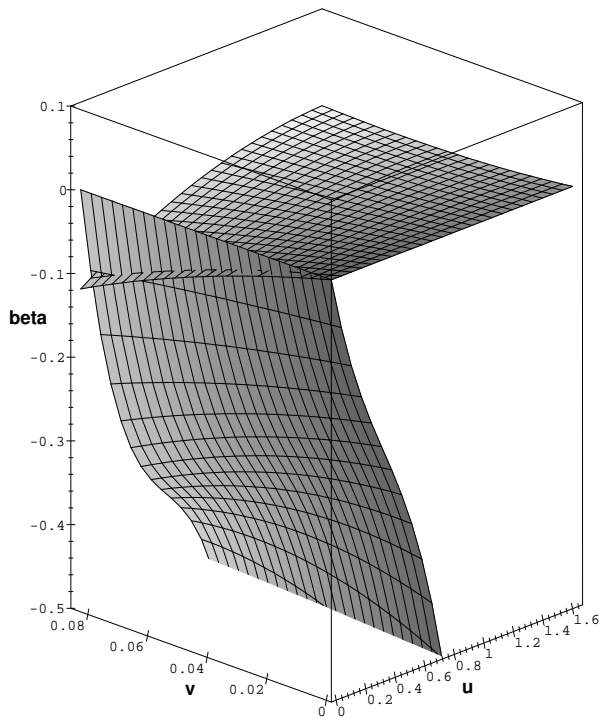}}
\end{picture}\\
\end{centering}
\caption{\label{fig4}
$\beta$-functions of the diluted Ising model
$\beta_u(u,v)$,
$\beta_v(u,v)$ in two-loop approximation for $d=3$.
Only the Gaussian fixed point $u^*=v^*=0$ survives.
}
\end{figure}

Looking for the solutions of the fixed point equations for functions
(\ref{19f}), (\ref{19g}) one can show that in the one-loop approximation,
in addition to the Gaussian fixed point $u^*=v^*=0$, there exist
two more solutions $u^*=2/3, \, v^*=0$ and  $u^*=0, \, v^*=-1/5$ and
the solution $u^*\neq 0, \, v^*\neq 0$ is absent\refnote{\cite{note5}}.
The fixed point with $v^*<0$ is beyond the region of parameters describing the
diluted magnet\refnote{\cite{note6}} and the pure model fixed point $u^*\neq0,
v^*=0$ appears to be unstable with respect to the $v$-coupling (we
propose to the reader to check this by looking at the stability
matrix $B_{ij}(u,v)$ (\ref{stab}) eigenvalues at the fixed points).
The corresponding plot of the functions $\beta_u$, $\beta_v$ in the
one-loop approximation is shown in the figure \ref{fig3}. Passing to
the two-loop approximation makes the result even worse: only the
Gaussian fixed point is present (see Fig. \ref{fig4}). Returning back
to the initial problem statement one should conclude that the obtained
picture corresponds to the absence of a second order
phase transition in a $d=3$ Ising model with a weak dilution as well as
without a dilution (the absence of the fixed point $u^*\neq 0, \, v^*= 0$).
Of course, this is contrary to the real situation. Let us further note
that the behaviour obtained for the $\beta$-functions of the model
(\ref{16a}) in the one- and two-loop approximations (Figs.
\ref{fig3}, \ref{fig4}) resembles those for the superconductor
case in the corresponding approximations (Figs. \ref{fig1},
\ref{fig2}).

\begin{figure}[ht]
\begin{centering}
\setlength{\unitlength}{1mm}
\begin{picture}(126,110)
\epsfxsize=126mm
\epsfysize=90mm
\put(10,0){\epsffile[22 20 822 582]{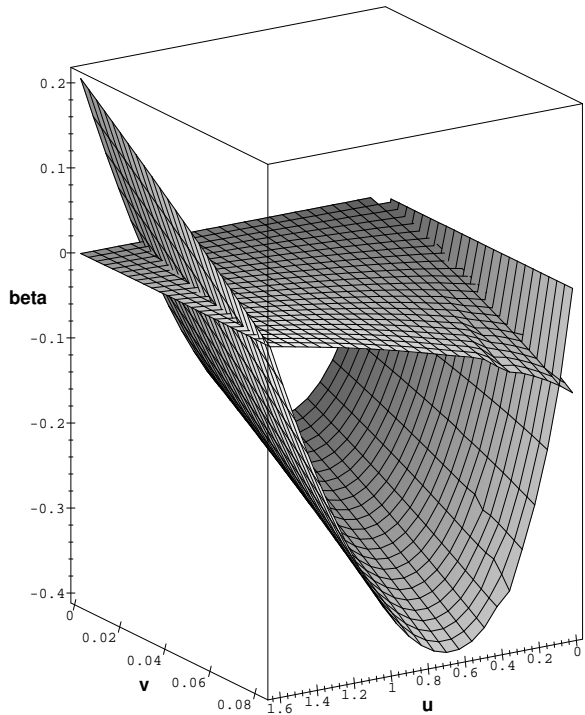}}
\end{picture}\\
\end{centering}
\caption{\label{fig5}
$\beta$-functions of the diluted Ising model
$\beta_u(u,v)$, $\beta_v(u,v)$ in two-loop approximation
for $d=3$ obtained by applying Pad\'e-Borel resummation technique.
Resummation restores the presence of the fixed point
$u^*\neq 0, \, v^*=0$ and results in the appearance of a new stable
fixed point  $u^*\neq 0, \, v^*\neq 0$.
}
\end{figure}

Applying, however, the resummation procedure to the series (\ref{19f}),
(\ref{19g}) in the two loop approximation one reconstitutes fixed
points ($u^*\neq 0, \, v^* = 0$), ($u^*= 0, \, v^*\neq 0$) and obtains
a new stable fixed point $u^*\neq 0, \, v^*\neq 0$ which governs a
second order phase transition in a weakly diluted Ising model.
The picture obtained appears to be stable with respect to successive
accounts of the higher order terms in the perturbation theory, when the
appropriate resummation technique is applied. As we have already
claimed above, these RG results are confirmed by other different
theoretical approaches and correspond to the experimentally observed
second order phase transition in the weakly diluted Ising magnet
with critical exponents differing from those of the pure case. In
Fig. \ref{fig5} we show the crossing of the $\beta_u(u,v)$ and
$\beta_v(u,v)$ surfaces for the resummed function. The
calculations have been  performed by means of the Pad\'e-Borel resummation
technique for the resolvent series (\ref{19e}) of two-loop functions
(\ref{19f}), (\ref{19g}) as described above\refnote{\cite{note7}}.
The Gaussian ($u^*=v^*=0$) and pure ($u^*= 1.3146, \, v^*=0$) fixed
points can be seen at the rear of the cube. The cross-section of
$u$ and $v$ planes in the picture are chosen to pass through the stable
fixed point ($u^*=1.6330, \, v^*= 0.0835$) corresponding to a new
critical behaviour.

The example we have considered above is a typical situation happening in a
$d=3$ RG theory: when considered without an appropriate resummation
technique, the RG analysis might not only give imprecise
values for critical exponents but also a qualitatively wrong answer
about the absence of a stable fixed point for a certain model, resulting
in the absence of a second order phase transition.
Now with this information in hand let us pass to the analysis of a
model of a superconductor described in the two-loop approximation by the
RG functions (\ref{19f}), (\ref{19g}).

\section{5 FIXED POINTS AND FLOWS IN THREE DIMENSIONS \label{V}}

We shall continue here by  considering the flow equations
(\ref{9}) directly for $d=3$.
We shall look for the solutions of the fixed point equations at $d=3$
paying attention to the possible asymptotic nature of the
corresponding series (\ref{10}), (\ref{11}).
Consider first the equation for the uncharged fixed point $U$.
Substituting $f^*=0$ into (\ref{11}) one obtains the
following expression for the function
$\beta_{u}^{U} \equiv \beta_{u}\;(u,f^*=0)$:
\begin{equation}
\beta_{u}^{U} = -u + \frac{n+8}{6} u^2 - \frac{3n+14}{12} u^3.
\label{20}
\end{equation}
Solving this polynomial for the fixed point one obtains
for the non-trivial $u^*>0$:
\begin{eqnarray}
&&u^{*U}= \frac{n+8}{3n+14} + \frac {\sqrt{n^2-20n-104}}{3n+14}
\label{21}
\end{eqnarray}
and immediately  the ``condition of the existence of a non-trivial
solution $u^{*U}$" qualitatively very similar to this, appearing
in the frame of the $\varepsilon$-ex\-pan\-si\-on technique, follows
(see\refnote{\cite{Halperin74,Kolnberger90}} and formula (\ref{13})  of the
present article as well): the solution exists only for certain values
of $n>n_c=24.3 \ $ !

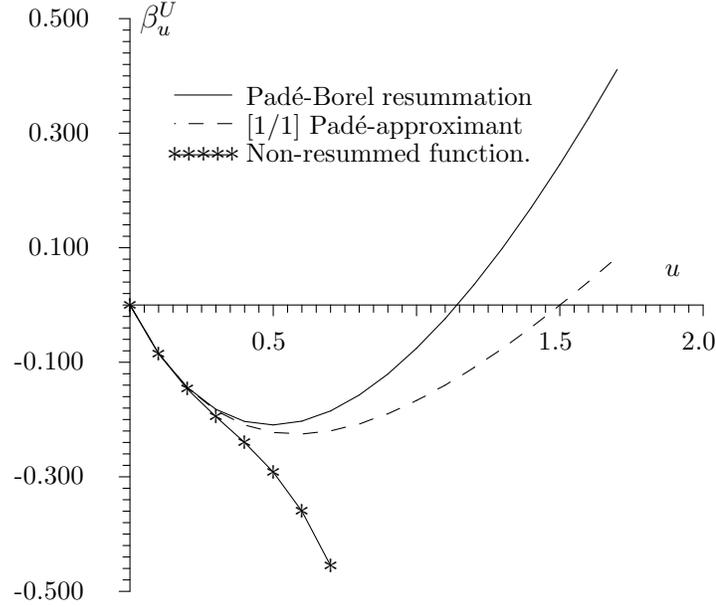
\begin{figure}[ht]
\begin{center}
\unitlength=0.5in
\begin{picture}(7.212,6.8)(0.788,0.7)
\input fig_u.pic
\end{picture}
\caption{
$\beta_u$-function of the uncharged model
$\beta_{u}^{U}$  at $d=3$, $n=2$.
\label{fig6}
}
\end{center}
\end{figure}

From Fig. \ref{fig6} one can see that
the function $\beta_{u}^{U}$ (\ref{20}) does not intersect the
$u$-axis for any non-zero value of $u$ for $n=2$.
In the $O(n)$-symmetric $\phi^4$-theory at $d=3$ this situation is
well known (see e.g.\refnote{\cite{Parisi,Holovatch93}}).  The
$\beta$-function calculated directly at $d=3$  does not
possess a stable zero for the realistic values of $n$, nevertheless with
in the three-loop order the presence of the stable fixed point is restored.
To avoid this artifact appearing in the two-loop calculation one
can either resum the series for the $\beta$-function or
construct the appropriate Pad\'e approximant\refnote{\cite{note8}}
in order to perform the analytical continuation of (\ref{20})
out of the domain of convergence (which is equal to zero for the
series in the right-hand side of (\ref{20})). Let us try both methods.
Representing (\ref{20}) in the form of a [1/1] Pad\'e approximant:
\begin{equation}
\beta_{u}^{U,Pad\acute{e}} = u  \frac{-1+A_u u}{1+ B_u u}
\label{22}
\end{equation}
one obtains:
\begin{equation}          \label{23}
A_u = \frac {n^2 + 7 n + 22}{6(n + 8)},
 \qquad
B_u = \frac {3 n + 14}{2(n +8)},
\end{equation}
\vspace{3mm}
and, solving the equation for the fixed point:
\begin{equation}
\beta_{u}^{U,Pad\acute{e}}(u^{*P,Pad\acute{e}}) = 0
\label{24}
\end{equation}
one obtains:
\begin{equation}
u^{*U,Pad\acute{e}} = \frac{6 (n + 8)}{n^2 + 7 n + 22}.
\label{25}
\end{equation}
So we have obtained a qualitatively different situation. The behavior of
the function $\beta_{u}^{U,Pad\acute{e}}(u)$  for $n=2$ is shown in
Fig. \ref{fig6} by the dashed curve. If one is interested in more
accurate values of $u^*$ some resummation has to be applied. Choosing
the Pad\'e-Borel resummation technique\refnote{\cite{49b}}
and following the scheme (\ref{17}) -- (\ref{19}) one obtains
for the resummed function $\beta_{u}^{U,Res}$\refnote{\cite{note7}}:
\begin{equation}
\beta_{u}^{U,Res} = u [2(1-A_u/B_u)(1-E(\frac{2}{uB_u})) \ - \ 1],
\label{26}
\end{equation}
the coefficients $A_u, \ B_u$ are given by
(\ref{23}),  $E(x) \ = \ x e^x E_1(x)$ , where
the function
$$
E_1(x) \ = \ e^{-x} \int\limits_0^{\infty} dt e^{-t} (x + t)^{-1}
$$
is connected with the exponential integral by the
relation\refnote{\cite{Abramovitz64}}:
$$
E_1(x \pm i 0) = - Ei(-x) \mp i \pi.
$$

The behavior of the function
$\beta_{u}^{U,Res}(u)$ is shown in Fig. \ref{fig6} by the solid
curve and the fixed point coordinate $u^{*U,Res}$ is obtained
solving the non-linear equation:
\begin{equation}
\beta_{u}^{U,Res}(u^{*U,Res}) = 0.
\label{27}
\end{equation}
The coordinates of the fixed point $u^{*U}$ obtained on
the basis of the Pad\'e approximation and the Pad\'e-Borel resummation
($u^{*U,Pad\acute{e}}, \quad u^{*U,Res}$) for different $n$ are given
in Table~\ref{tab1}.

\begin{table}
\begin{center}
\parbox{13cm}{
\caption{The fixed point U coordinate $u^{*U}$ as a function of $n$.
         $u^{*U,Pad\acute{e}}$: obtained on the basis of the [1/1] Pad\'e
         approximant;
         $u^{*U,Res}$: obtained by the Pad\'e-Borel resummation.}
\label{tab1}}
\begin{tabular}{lllllllll}
\\
\hline \hline
$n$ & 1 & 2 & 3 & 4 & 5 & 6 & 7 & 8 \\
\hline \\
$u^{*U,Pad\acute{e}}$  &
1.800 & 1.500 & 1.269 & 1.091 &  0.951 &  0.840 &  0.750 &  0.676 \\
\\
$u^{*U,Res}$ & 1.315 & 1.142 & 1.002 &  0.888 &  0.794 &  0.717 &
0.652 &  0.597 \\
\hline \hline
\end{tabular}
\end{center}
\end{table}

From this analysis we conclude that: in the $d=3$ theory the Pad\'e
approximants (as an analytical continuation of $\beta$-functions)
may change the picture qualitatively and lead to values of fixed
points comparable to those obtained by the Pad\'e-Borel resummation
technique.

\begin{figure}[ht]
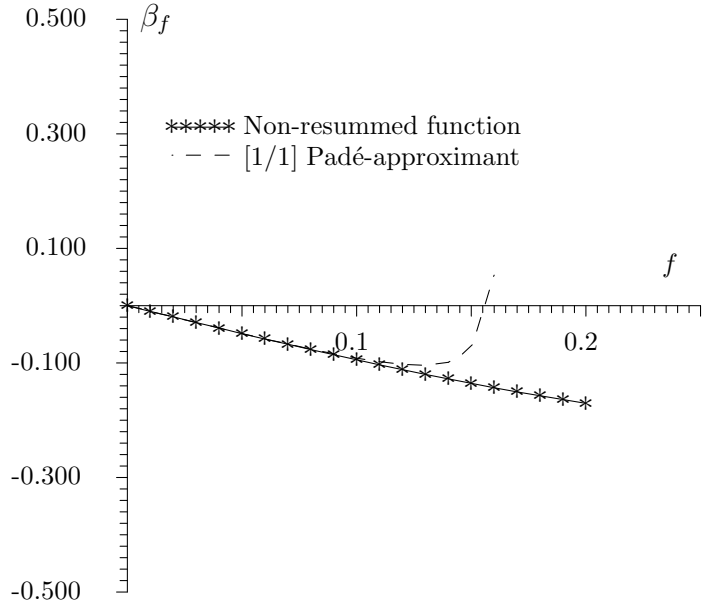

\begin{center}
\input fig_f.pic
\end{center}
\caption{
$\beta_{f}$-function at $d=3, n=2$.
\label{fig7}
}
\end{figure}

Consider now the equation for the charged fixed point $C$ applying the
above considerations to $\beta_f$, for which the expression at $d=3$
reads (\ref{10}):
\begin{equation}
 \beta_f = -f + \frac{n}{6} f^2 + n f^3. \label{28}
\end{equation}
The behavior of $\beta_f$ as a function of $f$ is shown in Fig.
\ref{fig7} by asterisks.  Note, however, that in this case the function
$\beta_{f}$ even without any resummation possesses a non-trivial zero
$f^{*M}$ (its value $f^{*C,Dir}$ is given in the 2nd row of Table
\ref{tab2}).  Representing (\ref{28}) in the form of the [1/1] Pad\'e
approximant:
\begin{equation}
\beta_{f}^{Pad\acute{e}} = f  \frac{-1+A_f f}{1+ B_f f}
\label{29}
\end{equation}
one has for $A_f$, $B_f$:
\begin{equation}
A_f = \frac {n + 36}{6}, \quad B_f = -6,
\label{af}
\end{equation}
and, solving the equation for the fixed point coordinate
$f^{*C,Pad\acute{e}}$:
\begin{equation}
\beta_{f}^{Pad\acute{e}}(f^{*C,Pad\acute{e}}) = 0
\label{30}
\end{equation}
one obtains:
\begin{equation}
f^{*C,Pad\acute{e}} = \frac{6}{n + 36}.
\label{31}
\end{equation}

\begin{table}
\begin{center}
\parbox{14cm}{
\caption {The fixed point C coordinate $f^{*C}$ as a
         function of $n$.  $f^{*C,Dir}$: obtained by a direct solution
         of the equation for the fixed point; $f^{*C,Pad\acute{e}}$:
         obtained on the basis of the [1/1] - Pad\'e approximant;
         $f^{*C,\varepsilon}$: the $\varepsilon$-expansion result with
         the linear accuracy in $\varepsilon$;
         $f^{*C,\varepsilon^2}$: the $\varepsilon$-expansion result
         with the square accuracy in $\varepsilon$.}
\label{tab2}}
\tabcolsep1.5mm
\begin{tabular}{lllllllll}
\\
\hline \hline
$n$ & 1 & 2 & 3 & 4 & 5 & 6 & 7 & 8 \\
\hline \\
$f^{*C,Dir}$ & 0.920  & 0.629  & 0.500  & 0.424  & 0.372  & 0.333  &
0.304 & 0.280  \\ \\
$f^{*C,Pad\acute{e}}$ & 0.162 & 0.158 & 0.154 & 0.150 & 0.146 & 0.143
& 0.140 & 0.136 \\ \\
$f^{*C,\varepsilon}$ & 6.000  & 3.000  & 2.000  & 1.500  & 1.200 &
1.000  &  0.857  & 0.750  \\ \\
$f^{*C,\varepsilon^2}$ & -210.000 & -51.000  & -22.000  & -12.000  &
-7.440  & -5.000  & -3.551  & -2.625  \\
\hline \hline
\end{tabular}
\end{center}
\end{table}

The function $\beta_{f}^{Pad\acute{e}}(f)$ is shown in Fig. \ref{fig7}
by the dashed line, the coordinate  $f^{*C,Pad\acute{e}}$ is given in
the 3rd row of Table \ref{tab2}.  But now the series (\ref{28}) is
not alternating and this results in the presence of a pole (at
$f= 1/6$) in the approximant (\ref{29}). Therefore, (\ref{29})
correctly represents the function $\beta_{f}(f)$ only for $f < 1/6$.
Let us note however that for all positive $n$ a fixed point exists
and its coordinate $f^{*M,Pad\acute{e}}$ lies within the limits $0 <
f^{*C,Pad\acute{e}} < 1/6$, where there are no poles in (\ref{29}).
Comparing this result with that obtained for the uncharged fixed
point one can note that the representation
of $\beta_{f}$ in the form of the Pad\'e approximant does not
qualitatively change the picture (a solution for $\beta_{f}(f) = 0$
exists at $d=3$ even without an analytical continuation) but results
in a decrease of the fixed point coordinate. Contrary to the
$\varepsilon$-expansion values (\ref{14}) there are no
borderline values of $n$ for the positivity of
$f^{*C}$.  Unfortunately, we cannot check this result by means of the
Pad\'e-Borel resummation technique.  The above mentioned presence of a
pole in the denominator of the Pad\'e approximant makes the corresponding
integral representation problematic\refnote{\cite{note9}}.
In order to find the $u$-coordinate of the fixed point C,
$u^{*C}$, we have to deal with a function of two variables,
$\beta_u(u,f)$,  represented by a rather short series
(\ref{11}). Another problem arises due to the fact that
the function $\beta_u(u,f)$ contains generating terms (i.e.
$\beta_u(u=0,f) \neq 0$). In order to perform some kind of an analytic
continuation of the function of two variables one can use
the Chisholm approximants (\ref{19c})\refnote{\cite{Chisholm73,Baker81}}.
But, the presence of generating terms
makes this choice rather ambiguous. The most reliable way in such
case is the representation of $\beta_u(u,f)$ in the form of a
resolvent series $B(u,f,\tau)$ (\ref{19e})\refnote{\cite{Watson74,Baker81}}
introducing an auxiliary variable $\tau$, which allows the separation
of contributions from different orders of the perturbation theory in the
coupling constants. The series for $B(u,f,\tau)$ then reads:
\begin{equation}
B(u,f,\tau) \equiv \beta_u(u\tau,f\tau) = \sum_{j\geq0} \ b_j \ \tau^j,
\label{32}
\end{equation}
with the usual notation applied for the coefficients $b_j$. Now one
considers (\ref{32}) as a series in the {\it single} variable $\tau$.
This series can be represented in the form of the Pad\'e approximant
$B^{Pad\acute{e}}(u,f,\tau)$ as an analytical continuation of the
function $B(u,f,\tau)$ for a general value of $\tau$. In particular
at $\tau=1$ the equality holds $B(u,f,\tau=1) = \beta_u(u,f)$  and the
approximant

\vspace{2mm}

$$
B^{Pad\acute{e}}(u,f,\tau=1) \equiv \beta_u^{Pad\acute{e}}(u,f)
$$

\vspace{5mm}

\noindent
represents the initial function $\beta_u(u,f)$. In our case the
expression for $B(u,f,\tau)$ reads:

\vspace{1mm}

\begin{equation}
B(u,f,\tau) =  \tau ( b_1 + b_2 \tau + b_3 \tau^2),
\label{33}
\end{equation}
where:
\begin{eqnarray}
\nonumber
b_1= -u,\qquad b_2 = \frac {n + 8}{6} u^2 - 6 u f + 18 f^2,
\end{eqnarray}
\begin{equation} \nonumber
b_3 = -\frac{3 n + 14}{12} u^3 + \frac{2 n + 10}{3} u^2 f
+ \frac{71 n + 174}{12} u f^2 - (7 n + 90) f^3.
\end{equation}

\begin{figure}[ht]
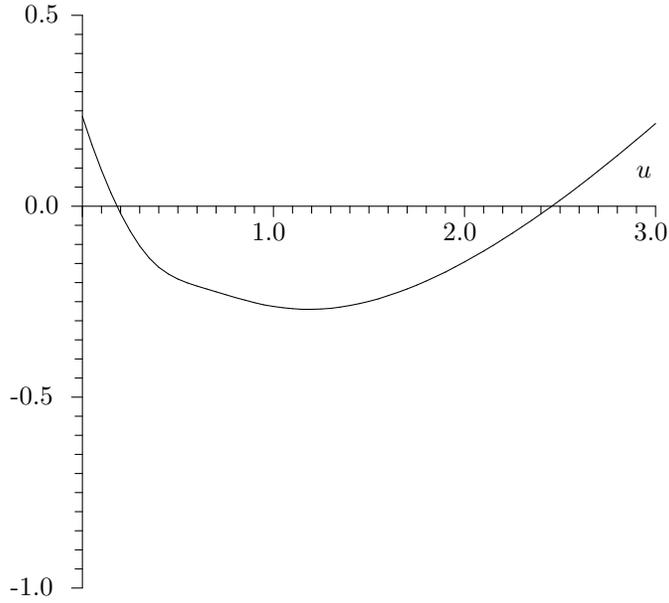

\begin{center}
\input fig_uf.pic
\end{center}
\caption{
Intersection of the function
$\beta_u^{Pad\acute{e}}(u,f)$ at $d=3$, $n=2$ with the plain
$f=f^{*C,Pad\acute{e}}$
in two-loop approximation.
\label{fig8}
}
\end{figure}

\noindent
Representing the expression in brackets in the right-hand side of
(\ref{33}) in the form of a [1/1] Pad\'e approximant we have:
\begin{equation}
B^{Pad\acute{e}}(u,f,\tau) = \tau \ b_1 \frac {1+A_{u,f}\tau}{1+B_{u,f}\tau},
\label{34}
\end{equation}
where
\begin{equation}
A_{u,f} = \frac{b_2}{b_1} - \frac{b_3}{b_2}, \quad
B_{u,f} = \frac{-b_3}{b_2}.
\label{auf}
\end{equation}

\begin{table}
\begin{center}
\parbox{12.5cm}{
\caption { The fixed point C coordinates
$u^{*C,Pad\acute{e}}$ obtained on the basis of the [1/1] Pad\'e
          app\-ro\-xi\-mant for the ``resolvent" series as a function
         of $n$.  C1 : the unstable fixed point; C2 : the stable fixed point.
         \label{tab3}}}
\begin{tabular}{lllllllll}
\\ \hline \hline
$n$ & 1 & 2 & 3 & 4 & 5 & 6 & 7 & 8 \\
\hline  \\
C1  & 0.184  & 0.181  & 0.179  & 0.177  & 0.175  & 0.175  & 0.176  &
0.179  \\ \\
C2 & 3.309 &  2.457 &  1.781 &  1.150 &  0.473 &  0.369 &
0.305 & 0.256 \\
\hline \hline
\end{tabular}
\end{center}
\end{table}

\noindent
Let us note here that the function $B(u,f,t)$
obtained in this way as the approximant for the function of two
variables
$\beta_u(u,f)$ obeys certain projection properties in the
single-variable case: substituting $f=0$ or $u=0$ into (\ref{34})
one obtains the [1/1]  Pad\'e approximant for $\beta_u^U(u)$ or the [0/1]
Pad\'e approximant for $\beta_u(u=0, f)$.
Finally the expression for $\beta_u(u,f)$ approximated in such way
reads:
\begin{equation}
\beta_u^{Pad\acute{e}}(u,f) =\ b_1 \frac {1+A_{u,f}}{1+B_{u,f}}.
\label{35}
\end{equation}
Substituting into the equation for the fixed point
$\beta_u(u^{*C},f^{*C})=0$ the value for the coordinate
$f^{*C}=f^{*C,Pad\acute{e}}$ (\ref{31}) one obtains
the non-linear equation for $u^{*C,Pad\acute{e}}$:
\vspace{2mm}
\begin{equation}
\beta_u^{Pad\acute{e}}(u,f=f^{*C,Pad\acute{e}} ) = 0.
\label{36}
\end{equation}

\begin{figure}[htbp]
\begin{centering}
\setlength{\unitlength}{1mm}
\begin{picture}(126,110)
\epsfxsize=126mm
\epsfysize=90mm
\put(10,0){\epsffile[22 20 822 582]{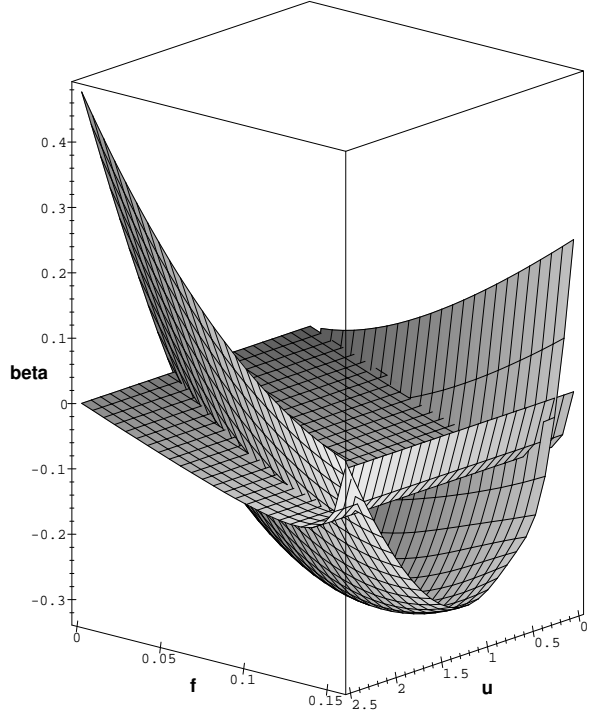}}
\end{picture}\\
\end{centering}
\caption{\label{fig9}
$\beta$-functions of the model of superconductor $\beta_u(u,f)$,
$\beta_f(u,f)$ in two-loop approximation for $d=3$, $n=2$ obtained by
Pad\'e analysis for the resolvent series. The stable "charged" fixed
point C2 with coordinates $u^*=2.457 , \, f^*= 0.158$
as well as the unstable fixed point C1  $u^*=0.181 , \, f^*= 0.158$
are seen on the front side of the cube. Gaussian ($u^*= f^*=0$) and
"uncharged" ($u^*=1.500, \,  f^*=0$) fixed points are located on the
rear of the cube.
}
\end{figure}

\noindent
Solving (\ref{36}) with respect to $u$ one obtains the values
$u^{*C,Pad\acute{e}}$ given in Table \ref{tab3}. The intersection of
the function $\beta_u^{Pad\acute{e}}(u,f)$ (\ref{35}) with the
plane $f=f^{*C,Pad\acute{e}}$ is shown for $n=2$ in Fig. \ref{fig8}.
The first fixed point (C1)  given in the 2nd row of Table
\ref{tab3} turns out to be unstable, while the fixed point C2 is
stable in the case $n=2$ we are predominantly interested in.

The resulting picture of $\beta$-functions surfaces is shown in the
Fig. \ref{fig9} . The Gaussian and uncharged fixed points may be
seen at the rear of the picture, whereas the intersection of the
$u$- and $f$-planes has been chosen in the picture to cross the stable
fixed point C2. The unstable fixed point C1 is seen as well.

The crossover to the asymptotic critical behavior is described by the solutions
 of
the flow equations (\ref{9}) with the initial
values of $u(\ell_0)$ and $f(\ell_0)$ at $\ell=\ell_0$\refnote{\cite{note10}}.
Substituting the analytical continuation of the $\beta$-functions in
the right-hand side of (\ref{9}), with the
Pad\'e approximants (\ref{29}), (\ref{35}) we get the following system
of differential equations:

\medskip

\begin{eqnarray}
l \frac{df}{dl} &=& f \frac{-1+A_f f}{1+ B_f f},
\label{37}
\\
l \frac {du}{dl} &=& - u \frac {1+A_{u,f}}{1+B_{u,f}},
\label{38}
\end{eqnarray}

\medskip

\noindent
where $A_f$, $B_f$ and $A_{u,f}$, $B_{u,f}$ are given by
(\ref{af}) and  (\ref{auf}), respectively.

\begin{figure}[htbp]
\begin{centering}
\setlength{\unitlength}{1mm}
\begin{picture}(126,90)
\epsfxsize=126mm
\epsfysize=90mm
\put(0,0){\epsffile[22 20 822 582]{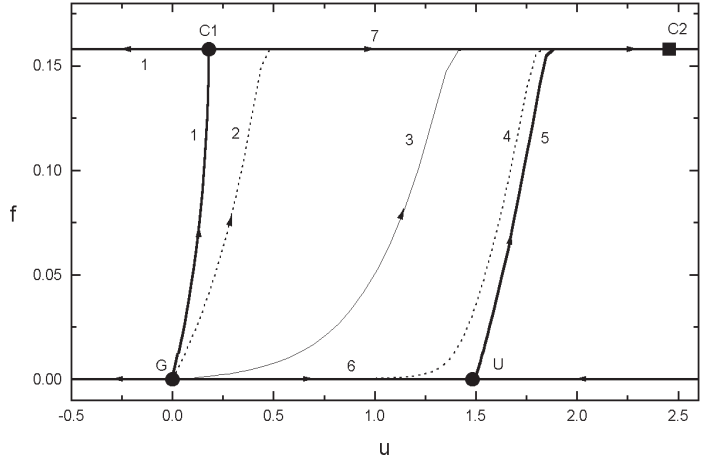}}
\end{picture}\\
\end{centering}
\caption{\label{fig10}
Flow lines for the case $n=2$, $d=3$ given
by equations (\protect\ref{37}), (\protect\ref{38}).
Fixed points G, U, C1 are unstable, fixed point C2 (shown by a box)
is a stable one (for further description see text).
}
\end{figure}

Solving equations (\ref{37}), (\ref{38}) numerically one obtains the
flow diagram shown in Fig. \ref{fig10} for the case $n=2$. The
space of the couplings is divided into several parts by separatrices
(thick lines in Fig. \ref{fig10}) connecting the fixed points. Besides
the Gaussian (G) there exist three fixed points, one corresponding to
the uncharged (U) and two others corresponding to the charged (C1, C2)
cases.  The fixed points G, C1 and U are unstable (solid circles in
Fig. \ref{fig10}) and the fixed point C2 is stable (shown as a
solid box in Fig. \ref{fig10}). Several different flow lines are shown
in Fig. \ref{fig10}. They can be compared with the corresponding flow
picture obtained by a direct solution of the flow equations for the
two-loop $\beta$-functions expressed by the third-order polynomials in
couplings $u$, $f$ (\ref{10}), (\ref{11}) (see Fig. 2a
in\refnote{\cite{Kolnberger90}}). There, one can see that no stable fixed point
exist and furthermore that the fixed point U is absent. Comparing Fig.
\ref{fig10} and Fig. 2b from\refnote{\cite{Kolnberger90}} one can see how an
analytical continuation of the $\beta$-functions (\ref{9}),
(\ref{10}), done only partly in\refnote{\cite{Kolnberger90}} and performed here
in the form of Pad\'e approximants restores the presence of the fixed
point U (unstable)
and leads to the appearance of a new stable fixed point C2 for the
charged model. The coordinates of the fixed points U, C1, C2
are given in the corresponding rows of Tables \ref{tab1},
\ref{tab2}, \ref{tab3} and for $n=2$ they are equal to:

\medskip

\begin{eqnarray*}
&&\makebox{U}:  u^{*}=1.500,  f^{*}= 0, \\
&&\makebox{C1}: u^{*}= 0.181,  f^{*}= 0.158, \\
&&\makebox{C2}: u^{*}= 2.457,  f^{*}= 0.158.
\end{eqnarray*}

\section{6 CRITICAL EXPONENTS \label{VI}}

Values of the critical exponents can be determined by the fixed point
values of the $\zeta$-functions defined on the basis of renormalizing
$Z$-factors (\ref{4}) - (\ref{7}) by:
\begin{equation}
\zeta_i = \mu \partial \ln Z_i/ \partial \mu,
\label{39}
\end{equation}
where the derivative is taken at fixed unrenormalized couplings.
The expressions for the
$\zeta$-functions related to the order parameter and the temperature
field renormalization  in the two-loop approximation
read\refnote{\cite{Kolnberger90}}: \begin{eqnarray}
\zeta_{\psi} &=& -3f + \frac{(n+2)}{72}u^2 + \frac{(11n+18)}{24}f^2,
\label{40}
\\ \nonumber
\zeta_{t} &=& \frac{-(n+2)}{6}u + \frac{(n+2)}{12}u^2 -
\\
&&\frac{2(n+2)}{3}uf - \frac{(5n+1)}{2}f^2,
\label{41}
\\
\zeta_A &=& \frac{n}{6}f+nf^2.
\label{41a}
\end{eqnarray}
If there exists a stable fixed point, the critical exponent of
the correlation length $\nu$, of the order
parameter susceptibility $\gamma$ and of the
specific heat $\alpha$ will be given by:
\begin{eqnarray}
\nu &=& (2- \zeta_{\nu}^{*})^{-1},
\label{42} \\
\gamma &=& (2- \zeta_{\nu}^{*})^{-1} (2- \zeta_{\psi}^{*}),
\label{43} \\
\alpha &=& (2- \zeta_{\nu}^{*})^{-1}
(\varepsilon- 2\zeta_{\nu}^{*}),
\label{44} \\
\eta &=&  \zeta_{\psi}^{*},
\label{44a}
\end{eqnarray}
where $\zeta_{\nu}=\zeta_{\psi} - \zeta_{t}$. The exponents
(\ref{42}) - (\ref{44a}) are related by the familiar scaling laws.
From the analysis
above it follows that the charged fixed point $C2$ is stable
and this results in values for exponents (\ref{42})--(\ref{44})
that are different from the values of the uncharged fixed point U,
i.e. they are not given by the $^4He$ values as it is sometimes stated
(see e.g.\refnote{\cite{Lobb87,Kiometzis94,Lawrie94}}).

Recently, an interesting consequence of the existence of a stable charged
fixed point ($C_2$) has been observed\refnote{\cite{Herbut96}}. According to
the charge renormalization (\ref{3}) the
$\beta_f$-function reads

\begin{equation} \label{zetaa}
\beta_f=f\ (\varepsilon -\zeta_A(f,u)).
\end{equation}
Thus at a fixed point with a nonzero $f^*$ the value of the
gauge field $\zeta$-function is given by
$\zeta_A^*=\varepsilon$ {\bf exactly}. This means that the penetration
depth $\lambda$ and the correlation length $\xi$ are proportional and
the temperature dependence follows a power law with the exponent
$\nu$\refnote{\cite{Herbut96}}. This is not the case, though, at the fixed
point with  $f^*=0$.
There we have $\zeta_A^*=0$ (each loop contribution to the $\zeta_A$-function
contains at least one $f$-factor) and the penetration depth behaves as
$\lambda\sim\xi^{(2-\varepsilon)/2}$. Thus one would have two different
critical length scales.

Now, trying to obtain numerical values of the critical exponents
on the basis of the values of fixed point $C2$ coordinates
 $f^{*C,Pad\acute{e}}$,
$u^{*C2,Pad\acute{e}}$ given in Tables~\ref{tab1}, \ref{tab2}, in
order to be self-consistent let us perform the same type of an analytical
continuation for the $\zeta$-functions series as those which
have applied to the  $\beta$-functions (\ref{10}),
(\ref{11}). So, after introducing the auxiliary variable $\tau$ let us
represent functions
(\ref{42}) - (\ref{44}) in the form of resolvent series in $\tau$
and then chose the [1/1] Pad\'e approximants for these series,
which at $\tau=1$ will give us the analytical continuation of the
series requested. Thus, the expression for the critical
exponent $\phi\,(\phi \equiv \{\nu, \gamma, \alpha)\}$ reads:
\begin{equation}
\phi =  a_{\phi}^{(0)} \ \frac {1 + A_{\phi}} {1 + B_{\phi}}.
\label{45}
\end{equation}
The expressions for the coefficients $A_{\phi}$, $B_{\phi}$ in
(\ref{45}) read:
\begin{equation}
A_{\phi}= a_{\phi}^{(1)} \ + \ B_{\phi} ;
\quad B_{\phi} = - a_{\phi}^{(2)}/a_{\phi}^{(1)},
\label{46}
\end{equation}
and $a_{\phi}^{(i)}$ are to be determined from the resolvent
series in $\tau$:
\begin{equation}
\phi = \sum_{i\geq0} a_{\phi}^{(i)} \ \tau^i |_{\tau=1}.
\label{47}
\end{equation}
Substituting (\ref{40}) and (\ref{41}) into (\ref{42}) -
(\ref{44}) and representing (\ref{42}) - (\ref{44}) in the
form of (\ref{47}) one finds:
\begin{eqnarray}
&&a_{\nu}^{(0)}= 1/2,
\nonumber \\
&&a_{\nu}^{(1)} =(n+2)/12 \ u - 3/2 \ f,
\nonumber \\
&&a_{\nu}^{(2)}= (n^2 - n - 6)/144 \ u^2 + (71n+138)/48 \ f^2 +
(n+2)/12 \ uf ,
\label{48} \\
&&a_{\gamma}^{(0)} = 1,
\nonumber \\
&&a_{\gamma}^{(1)}=(n+2)/12 \ u,
\nonumber \\
&&a_{\gamma}^{(2)}=(n^2 - 2n - 8)/144 \ u^2 + (5n+1)/4 \ f^2 +
5(n+2)/24 \ uf ,
\label{49} \\
&&a_{\alpha}^{(0)}=1,
\nonumber \\
&&a_{\alpha}^{(1)}=-3(n+2)/12 \ u \ + \ 9/2 \ f,
\nonumber \\
&&a_{\alpha}^{(2)}= (-3n^2+3n+18)/144 \ u^2 - (71n+138)/16 \ f^2  -
\nonumber \\
&& (n+2)/4 \ uf .
\label{50}
\end{eqnarray}
Now considering the case $n=2$ and substituting coordinates of the
fixed point $C2$  ($f^{*C,Pad\acute{e}} = .158$),
$u^{*C2,Pad\acute{e}} = 2.457$ (see Tables \ref{tab1}, \ref{tab2})
into (\ref{48}) - (\ref{50}) one obtains for the critical
exponents (\ref{42}) - (\ref{44a})\refnote{\cite{note11}}:

\smallskip

\begin{eqnarray}
&&\nu = 0.86, \quad \gamma=  1.88,
\\ \nonumber
&&\alpha = -1.14, \quad \eta = -0.19.
\label{exppade}
\end{eqnarray}

\smallskip

The application of the Pad\'e approximants for  the analytical
continuation of the functions may result in the appearance of poles
in these functions.  If the pole is located in a region of the expansion
parameters which has no physical meaning, e.g. a negative coupling $u$ or $f$,
the analysis is not complicated. This is the case for  the $\beta$-functions
in the region of couplings less than the fixed point values.
For the $\zeta$-functions, however, considering the non-asymptotic
behavior (and thus being far from the stable fixed point) one
passes through a region of couplings where the Pad\'e approximation
for the $\zeta$-functions becomes ambiguous which results in the
appearance of a pole. Therefore, studying the crossover behavior
in the next subsection  we will
still keep the polynomial representation for $\zeta$-functions instead
of the Pad\'e approximants. Then for the asymptotic values of
critical exponents one gets:

\smallskip

\begin{eqnarray}
&&\nu = 0.77, \quad \gamma=  1.62,
\\ \nonumber
&& \alpha = -0.31,
\quad \eta = -0.10.
\label{exp}
\end{eqnarray}

\smallskip

\noindent
The comparison of the values
(\ref{exppade}) and (\ref{exp}) shows a numerical
difference of 15\% in $\nu$ and  $\gamma$ and a considerable
increase of $\alpha$ but there is no qualitative change (e.g. the sign of
 the
specific heat exponent remains the same). This should be compared with
values given by others, $\nu = 0.53$ and $\eta=-0.70$\refnote{\cite{Herbut96}}
and $\eta=-0.38$\refnote{\cite{Radzihovsky95}}.  Since for the
conventional superconductors the experimentally
accessible re\-gi\-me lies in the precritical region further away from
$T_c$, let us now  discuss some non-asymptotic quantities such as
effective exponents and amplitude ratios.

\section{7 AMPLITUDE RATIO FOR THE SPECIFIC HEAT
\label{VII}}

One of the most interesting measurable quantities is the specific
heat. Near a second order phase transition, asymptotically it follows a
power law
\bigskip

\begin{equation}
C^{\pm}_0=\frac{A^{\pm}}{\alpha}|t|^{-\alpha}+\mbox{const.}
\end{equation}

\smallskip
\noindent
where $\pm$ indicates the specific heat $C$ and its non universal
amplitude $A$ above and below $T_c$. The amplitude ratio
$A^+/A^-$ found from the ratio $C^+(t^+)_0/C^-(t^-)_0$ after
subtracting the non singular background value
constitutes a universal quantity at $T_c$ depending only on
the dimension and the number of components of the order parameter.

The calculation of this ratio can be extended to the
non asymptotic region\refnote{\cite{DOPRL,DOGARR}}
resulting in a temperature
dependent measurable quantity. This also tests the description
of the nonasymptotic behavior by a certain flow in the interaction
space of the Hamiltonians, discussed for the effective
exponents. The starting point in the calculation is the
renormalization group equation for the specific heat $C^\pm$

\bigskip
\begin{equation}
\left[\mu\frac{\partial}{\partial \mu}+\beta_u\frac{\partial}
{\partial u}+\beta_f\frac{\partial}{\partial f}+\zeta_\nu
\left(2+t\frac{\partial}{\partial t}\right)\right]
C^\pm(t,u,f,\mu)=\mu^{-\varepsilon}B(u,f),
\end{equation}

\smallskip
\noindent
where the inhomogeneity $B$ comes from the additive
renormalization. The formal solution reads

\bigskip
\begin{eqnarray}
C^\pm(t,u,f,\mu)=\mu^{-\varepsilon}\exp\left[-
\int\limits_1^l(\varepsilon-2\zeta_nu(x)\frac{dx}{x}\right]
\;\;\;\;\;\;\;\;\nonumber \\
\times\left\{F^\pm(l)-\int\limits_1^l\frac{dy}{y}B(y)\exp\left[-
\int\limits_l^y(\varepsilon-2\zeta_\nu(x))\frac{dx}{x}\right]
\right\}.
\end{eqnarray}

\smallskip
\noindent
The amplitude ratio is most easily calculated by choosing
the same value for the flow parameter both above and below $T_c$,
which means $t^+=-2t^-$\refnote{\cite{DOZEIT}}.
We then recover the asymptotic expression found in\refnote{\cite{DOGARR}}

\bigskip
\begin{equation}
\frac{A^+}{A^-}=2^\alpha\frac{B\nu+F^+\alpha}{B\nu+F^-\alpha},
\end{equation}

\smallskip
\noindent
where the functions $B$ and $F^\pm$ are taken at the fixed point.
For this function we use the lowest order result known from
the $\Psi^4$ theory neglecting the coupling to the gauge field;
$B=2n$, $F^+=-n$ and $F^-=12/u^*-4$. Then we have for $n=2$

\bigskip
\begin{equation}
\frac{A^+}{A^-}=2^\alpha\frac{2\nu-\alpha}{2\nu-2\alpha+
6\alpha/u^*}.
\end{equation}

\smallskip
\noindent
In the Table~\ref{tab4} we have collected the values obtained for the
different fixed points.
It is interesting to note the reasonable estimate for this ratio
presented in\refnote{\cite{SINGSAAS}} for the superfluid phase transition.
The authors have found a value of $A^+/A^-=1.067$, which is surprisingly
near the value obtained in the stable charged fixed point using
both calculation schemes, although the exponents are very different.
We have not, however, taken into account changes in the scaling functions
due to the coupling $f$.

\begin{table}
\begin{center}
\parbox{10cm}{\caption{\label{ratio} Asymptotic values for the
specific heat amplitude ratio at various fixed points. The exponents
in the first two lines correspond to the procedure leading to
(\protect\ref{exppade}). The third and the fourth lines correspond to
the procedure leading to (\protect\ref{exp}).
\label{tab4}
}}
\begin{tabular}{ccccc} \\
\hline \hline
F.P. & $A^+/A^-$ & $\alpha$ & $\nu$ & $u^*$ \\ \hline \\
U    &   1.81    &  -0.33   &  0.72      &   1.500  \\ \\
C2   &   1.07    &  -1.14   &  0.86      &   2.457   \\ \\
U    &   0.78    &   0.15   &  0.62      &   1.500   \\ \\
C2   &   1.06    &  -0.31   &  0.77      &   2.457   \\ \hline \hline
\end{tabular}
\end{center}
\end{table}

In comparison with experimental results\refnote{\cite{Salamon88}}, the
amplitude ratio of the Gaussian $n$-vector model without a coupling to the
gauge field $A^+/A^-=n/2^{3/2}$ has been used since the dimension of the
order parameter is unclear. Later on this expression for the
amplitude ratio has been calculated for cases other than the isotropic
symmetry. This leads to a dependence of the ratio on the higher order
couplings\refnote{\cite{ANNETT}}.

\section{8 EFFECTIVE EXPONENTS
\label{VIII}}

Effective exponents are usually defined by the
logarithmic temperature derivatives of the corresponding correlation
functions (see e.g.\refnote{\cite{Chen78,Riedel74}}). They can be found from
the solutions of the renormalization group equation for the
renormalized vertex functions.  The effective exponents contain two
contributions, one from the corresponding $\zeta$-functions now taken
at values of $u(\ell)$, $f(\ell)$ of the particular flow curve considered
(``the exponent part"), and one from the change of the corresponding
scaling function (``the amplitude part").  For the analysis below,
we neglect the latter contributions since we expect
them to be smaller than the differences for the fixed point values of
the exponents coming from the different treatments discussed before.
Thus we have:

\begin{eqnarray} \nu &=& (2- \zeta_{\nu}(\ell))^{-1},
\label{51}
\\
\gamma &=& (2- \zeta_{\nu}(\ell))^{-1} (2-
\zeta_{\psi}(\ell)), \label{52}
\\
\alpha &=& (2-
\zeta_{\nu}(\ell))^{-1} (\varepsilon- 2\zeta_{\nu}(\ell)).
\label{53}
\end{eqnarray}
The flow parameter $\ell$ can be related to the
relative temperature distance $T_c$ by the matching condition
$t(\ell)=(\xi_0^{-1}\ell)^2$, with $\xi_0$ the amplitude of the
correlation length.

\begin{figure}[htbp]
\begin{centering}
\setlength{\unitlength}{1mm}
\begin{picture}(126,90)
\epsfxsize=126mm
\epsfysize=90mm
\put(0,0){\epsffile[22 20 822 582]{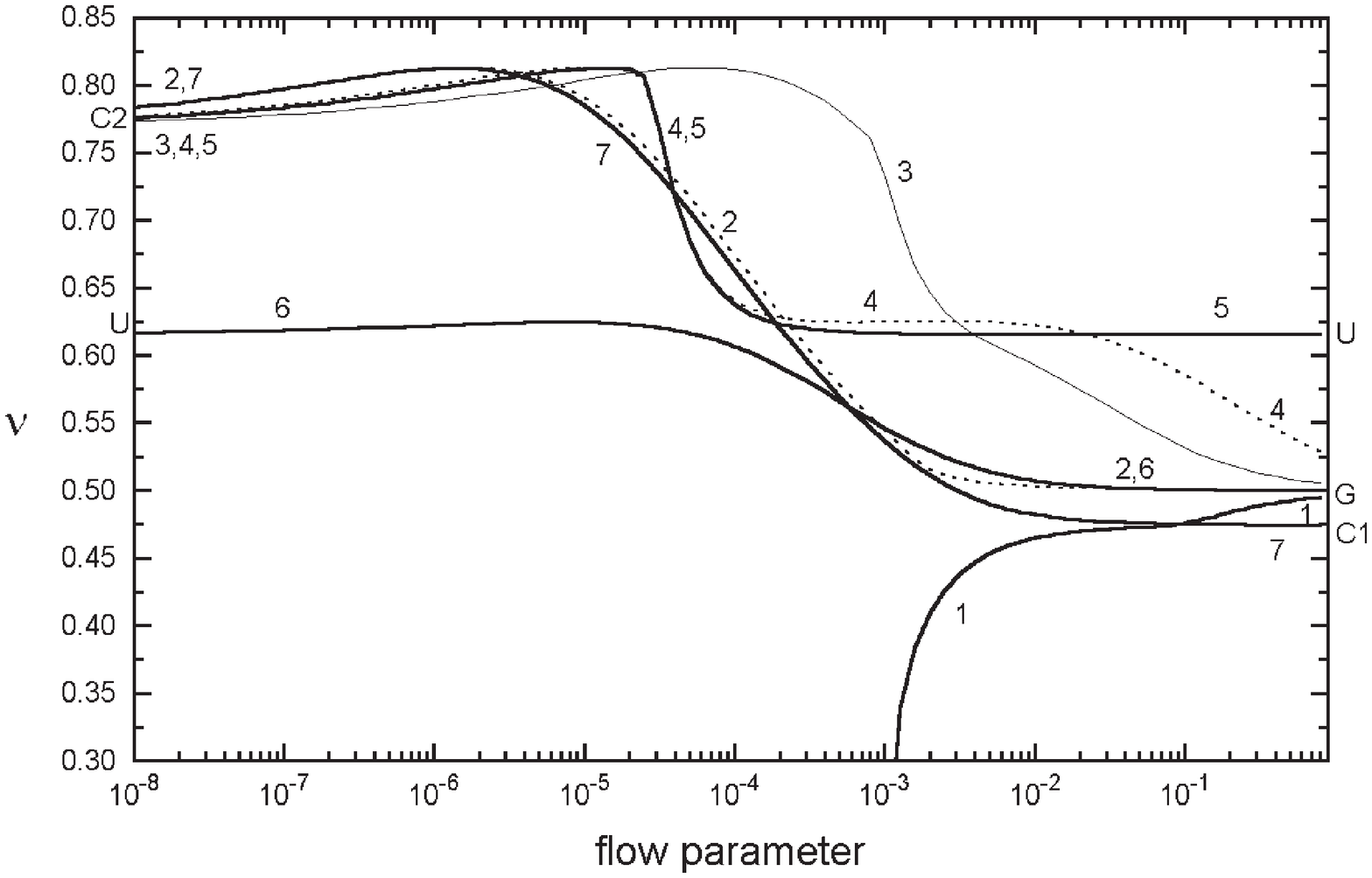}}
\end{picture}\\
\end{centering}
\caption{ \label{fig11}
Effective exponent $\nu$ for the flows
shown in Fig.\protect\ref{fig10}
(for further description see text).
}
\end{figure}

\begin{figure}[htbp]
\begin{centering}
\setlength{\unitlength}{1mm}
\begin{picture}(126,90)
\epsfxsize=126mm
\epsfysize=90mm
\put(0,0){\epsffile[22 20 822 582]{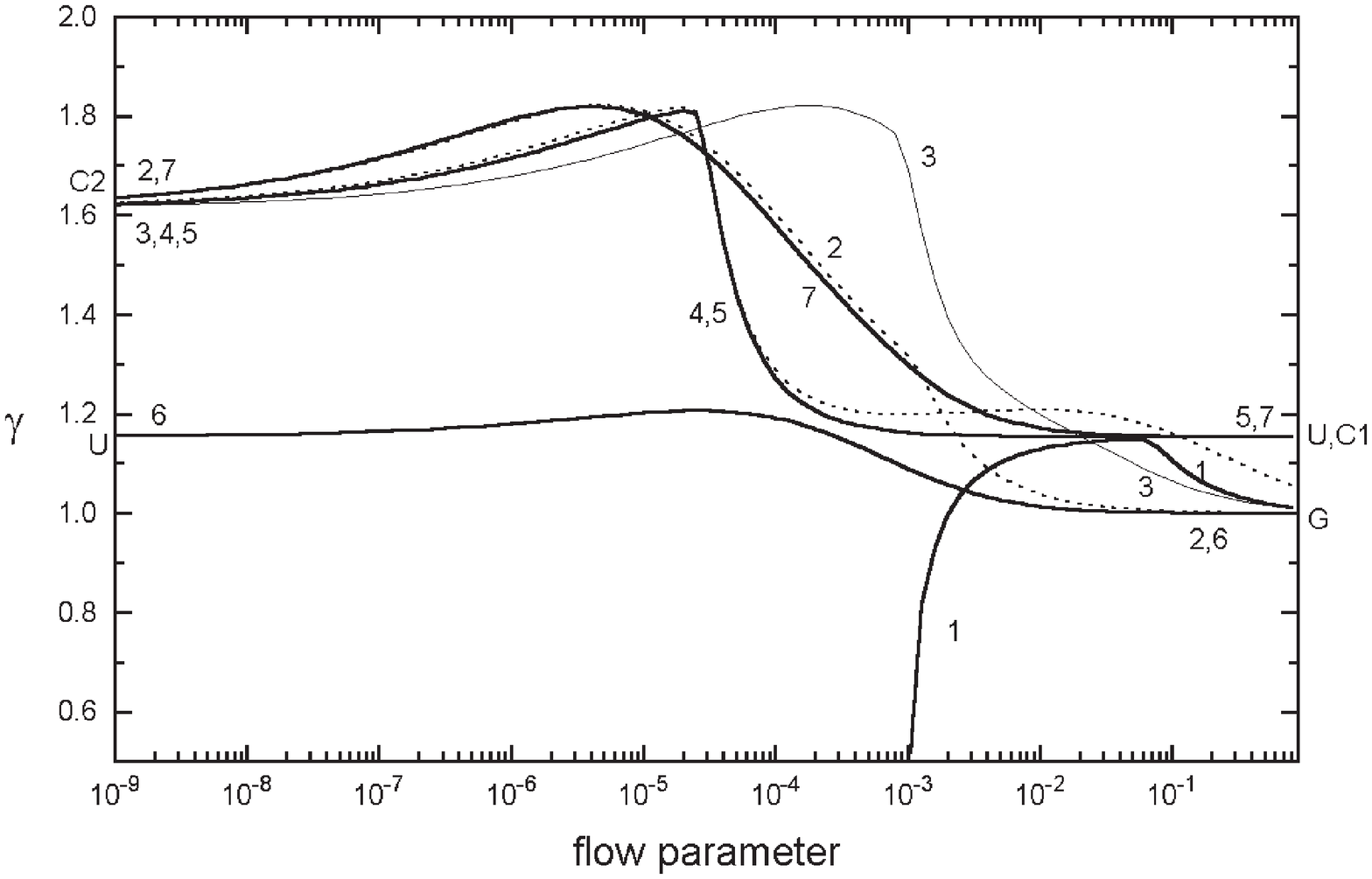}}
\end{picture}\\
\end{centering}
\caption{ \label{fig12}
Effective exponent $\gamma$ for the flows shown in
Fig.\protect\ref{fig10} (for further description see text).
}
\end{figure}

\begin{figure}[htbp]
\begin{centering}
\setlength{\unitlength}{1mm}
\begin{picture}(126,90)
\epsfxsize=126mm
\epsfysize=90mm
\put(0,0){\epsffile[22 20 822 582]{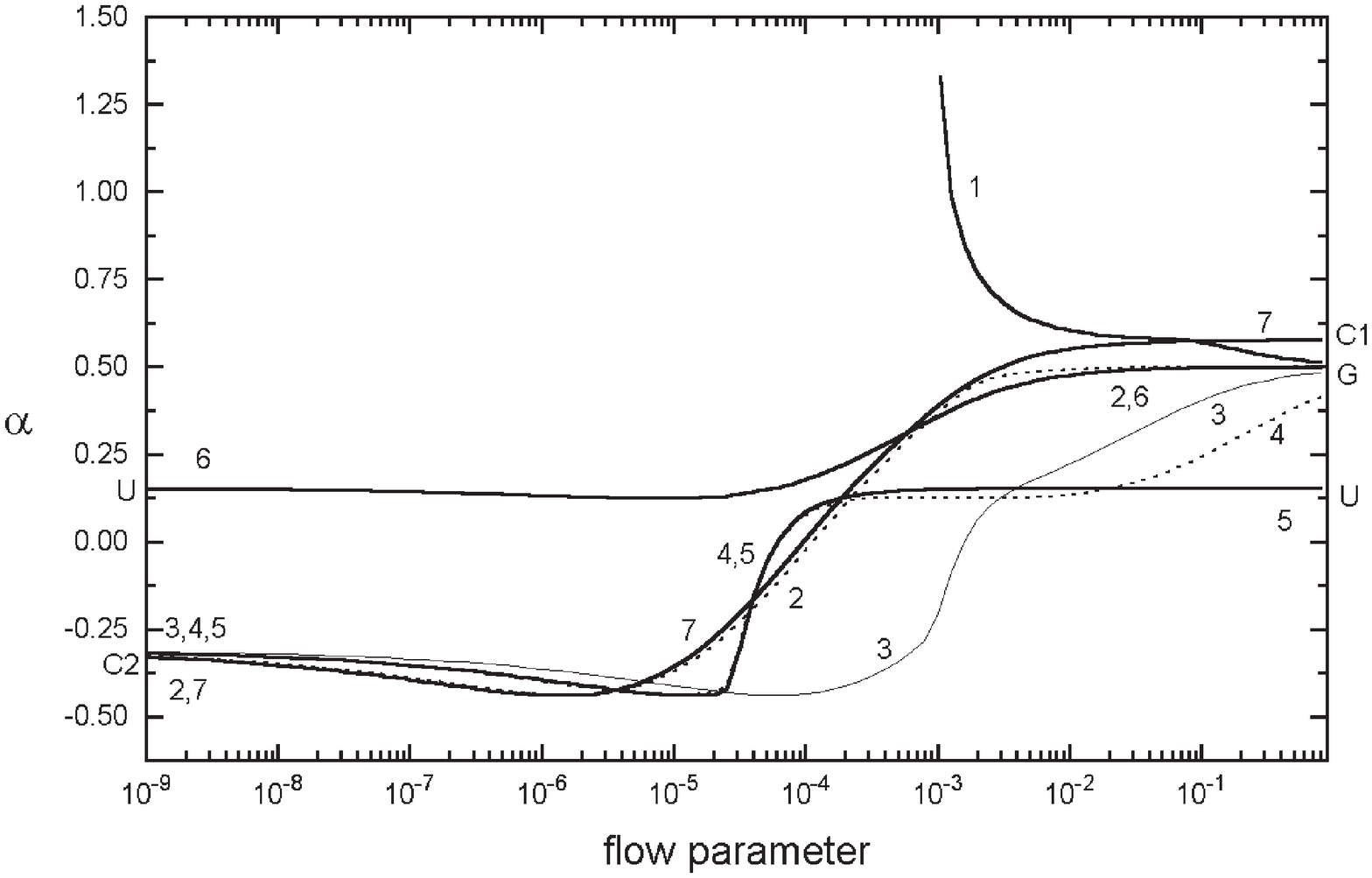}}
\end{picture}\\
\end{centering}
\caption{ \label{fig12a}
Effective exponent $\alpha$ for the flows
shown in Fig.\protect\ref{fig10}
(for further description see text).
}
\end{figure}

We have computed these effective exponents, see Fig. \ref{fig11} -
Fig. \ref{fig12a}, along the flow lines shown in Fig. \ref{fig10}
by inserting\refnote{\cite{note12}}
values of the couplings $u(\ell)$ and $f(\ell)$  into Eqs.
(\ref{51}) - (\ref{53}). For the separatrix 1 we have started with
initial conditions leading to a flow, which does not coincide with the fixed
point C1 but slightly misses it although the flow curve does not differ
from the separatrix within the thickness of the lines shown in
Fig. \ref{fig10}. For the curve number 4, we start somewhat further
away from the Gaussian fixed point G leading to the initial values of
the effective exponents between their Gaussian values and their values
for the uncharged fixed point U. Note that when the values of the effective
exponent $\gamma$ for the uncharged fixed point U and the charged
fixed point C1 are the same within the accuracy given
by the scale of the figure.

Note that when the coupling $f$ to the gauge field fluctuations is small
(i.e. for extreme type-II superconductors) the RG flow passes very closely
to the uncharged fixed point $U$ and the effective exponents, and
within some region of temperatures, they coincide with those of the
uncharged superfluid liquid. In this region the effective Hamiltonian
(\ref{2}) may be considered as that of a superconductor in a constant
magnetic field neglecting magnetic field fluctuations. Recently for
such a model it has been shown that near the zero-field critical point the
singular part of the free energy scales as $F_{sing}\simeq
|t|^{2-\alpha}{\cal F}(B|t|^{-2\nu})$ with $\nu$ being the coherence
length exponent\refnote{\cite{Lawrie97}}.

\section{9 CONCLUSIONS \label{IX}}

Does the above account give a definite conclusion about the order of
the phase transition occurring in a model of the superconductor minimally
coupled to the gauge field?  First of all one should keep in mind
that such an answer may be given in an analytical
theory framework only by obtaining an exact result or a rigorous proof.
Here, the problem has been treated by a perturbation theory approach and the
account of the influence of fluctuations on the order of the phase
transition is studied within the field theoretical RG technique. We
have shown that remaining within this approach one may get an answer about
the second order phase transition occurring in the above mentioned
model.

The main point discussed in this context
is whether the equations for $\beta$-functions possess a stable fixed
point or not. The absence of the stable fixed point is often
interpreted as a change of the order of the phase transition (caused by
the presence of the magnetic field fluctuations) and the evidence of the
fluctuation-induced first-order phase transition.  This change
of the order of the phase transition (being of the second-order in the
absence of the coupling to the gauge), however, is confirmed only by the
perturbation theory calculations in low orders\refnote{\cite{Halperin74}},
(see\refnote{\cite{Kolnberger90}} and the references therein as well).

Applying a simple Pad\'e analysis to the
series under discussion\refnote{\cite{note12a}}
we have shown how one can recover a stable fixed point in the RG
equations. In the case of one coupling, such an approach gives a
qualitatively correct picture of the phase transition and restores the
presence of a stable fixed point (\refnote{\cite{Parisi}}, see formulas
(\ref{21}), (\ref{25}) of this article as well).
The same occurs in the case of two
couplings: for $n=2$ the ``uncharged" fixed point U
(having coordinates $f^{*U,Pad\acute{e}} = .158$,
$u^{*U,Pad\acute{e}} = 2.457$) appears to be stable, which leads to a
new set of critical exponents.  We should note, however, that the pair
correlation function critical exponent $\eta$, calculated by familiar
scaling relations on the basis of sets of values (\ref{exppade})
or (\ref{exp}), remains negative which agrees with the result
of\refnote{\cite{Radzihovsky95,Herbut96,Bergerhoff96}}. Being calculated only
in a two-loop approximation with the application of the Pad\'e analysis, these
values for the critical exponents are to be considered as preliminary
ones.  The main point we wish to make is that within the framework of
the renormalization group analysis for the superconductor model there
remains the possibility of a second-order phase transition
characterized by a set of critical exponents differing from those of
$^4He$.  Another important task could be to calculate the nonasymptotic
specific heat in order to compare with experiments within the region
of the crossover to the background.

\section{ACKNOWLEDGEMENTS}

We acknowledge the useful discussions with A.M.J. Schakel,
D.I. Uzunov, A.E. Filippov. We are grateful to O.J. Poole who read the
manuscript and made suggestions for improving the text.


\begin{numbibliography}

\bibitem{Wilson71}
K.~G.~Wilson,  {\it Phys. Rev.} B 4:3174 (1971); ibid
4:3184 (1971).

\bibitem{Bogoliubov59}
N.~N.~Bogoliubov, D.~V.~Shirkov.
"Introduction to the Theory of Quantized Fields,"
Wiley \& Sons, New York (1959).

\bibitem{Amit84} D.~J. Amit. "Field Theory, the Renormalization
Group, and Critical Phenomena," World Scientific, Singapore (1984).

\bibitem{LeBellac91} M.~Le~Bellac. "Quantum and Statistical Field
Theory," Claredon Press, Oxford (1991).

\bibitem{Zinn96} J.~Zinn--Justin. "Quantum Field Theory and
Critical Phenomena," Oxford University Press, Oxford (1996).

\bibitem{Lipa96} J.~A.~Lipa, D.~R.~Swanson, J.~A.~Nissen,
T.~C.~P.~Chui, U.~E.~Israelsson, {\it Phys. Rev. Lett.} 76:944
(1996).

\bibitem{Baker78} G.~A.~Baker~Jr., B.~G.~Nickel, D.~I.~Meiron, {\it Phys.
Rev.} B 17:1365 (1978).

\bibitem{LeGuillou80} J.~C.~Le~Guillou, J.~Zinn-Justin, {\it Phys. Rev.} B
21:3976 (1980).


\bibitem{Bagnuls85} C.~Bagnuls, C.~Bervillier, {\it Phys. Rev.} B
32:7209 (1985).

\bibitem{Bagnuls87} C.~Bagnuls, C.~Bervillier, D.~I.~Meiron,
B.~G.~Nickel, {\it Phys. Rev.} B 35:3585 (1987).

\bibitem{Schloms87} R.~Schloms, V.~Dohm, {\it Europhys. Lett.} 3:413 (1987).

\bibitem{Schloms89} R.~Schloms, V.~Dohm, {\it Nucl. Phys.} B 328:639 (1989).

\bibitem{Schloms90} R.~Schloms, V.~Dohm, {\it Phys. Rev.} B 42:6142 (1990).

\bibitem{Vakarchuk78} I.~A.~Vakarchuk, {\it Theor. Math. Phys.} (Moscow)
36:122 (1978).

\bibitem{Halperin74} B.~I.~Halperin, T.~C.~Lubensky, S.~Ma, {\it Phys. Rev.
Lett.} 32:292 (1974).

\bibitem{Lobb87} J.~Lobb,  {\it Phys. Rev.} B 36:3930 (1987).

\bibitem{Inderhees88}
S.~E.~Inderhees, M.~B.~Salamon, N.~Goldenfeld, J.~P.~Rice,
B.~G.~Pazol, D.~M. Ginsberg, J.~Z.~Liu, G.~W.~Crabtree,
{\it Phys. Rev. Lett.} 60:1178 (1988).

\bibitem{Salamon88} M.~B.~Salamon, S.~E.~Inderhees, J.~P.~Rice,
B.~G.~Pazol, D.~M.~Ginsberg, N.~Goldenfeld, {\it Phys. Rev.} B 38:885 (1988).

\bibitem{Inderhees91}
S.~E.~Inderhees, M.~B.~Salamon, J.~P.~Rice, D.~M.~Ginsberg,
{\it Phys. Rev. Lett.} 66:232 (1991).

\bibitem{Regan91}
S.~Regan, A.~J.~Lowe, M.~A.~Howson, {\it J. Phys.: Condens. Matter} 3:9245
(1991).

\bibitem{Mozurkewich92} G.~Mozurkewich, M.~B.~Salamon, {\it Phys. Rev.} B
46:11914 (1992).

\bibitem{Salamon93}
M.~B.~Salamon, J.~Shi,  N.~Overend, M.~A.~Howson, {\it Phys. Rev.} B 47: 5520
(1993).

\bibitem{Junod93} A.~Junod, E.~Bonjour, R.~Calemczuk, J.~Y.~Henry,
J.~Muller,  G.~Triscone, J.~C.~Vallier, {\it Physica} C 211:304
(1993).

\bibitem{Overend94} N.~Overend, M.~A.~Howson, I.~D.~Lawrie, {\it Phys. Rev.
Lett.} 72:3238 (1994).

\bibitem{Lawrie94} I.~D.~Lawrie, {\it Phys. Rev.} B 50:9456 (1994).

\bibitem{Chen78} J.~H.~Chen, T.~C.~Lubensky, D.~R.~Nelson, {\it Phys.
Rev.} B 17:4274 (1978).

\bibitem{LeGuillou77} J.~C. Le~Guillou, E.~Br\'ezin, J.~Zinn-Justin,
{\it Phys. Rev.} D 15:1544 (1977).

\bibitem{Lipatov77} L.~N.~Lipatov, {\it Sov. Phys. JETP.} 45:216
(1977).

\bibitem{Brezin78} E.~Brezin, G.~Parisi, {\it J. Stat. Phys.} 19:269
(1978).

\bibitem{Kolnberger90} S.~Kolnberger, R.~Folk, {\it Phys. Rev.} B 41:4083 (1990)

\bibitem{Folk96} R.~Folk, Yu.~Holovatch, {\it J. Phys. A: Math. Gen.}
29:3409 (1996).

\bibitem{Folk97} R.~Folk, Yu.~Holovatch, {\it J. Phys. Stud.} (Lviv)
1:343 (1997).

\bibitem{Meyer96} H. Meyer-Ortmanns, {\it Rev. Mod. Phys.} 68:473 (1996)

\bibitem{note1} Recall that for the given Landau-Ginsburg
parameter (ratio of the penetration depth to the coherence length) $k$
a superconductor with $k < 1/ \sqrt 2$ is called  type-I and the one
with $k > 1/ \sqrt 2$ is called  type-II.

\bibitem{Wilson72} K.~G.~Wilson, M.~E.~Fisher, {\it Phys. Rev. Lett.}
28:240 (1972).

\bibitem{Filippov94} A.~E.~Filippov, A.~V.~Radievsky, A.~S.~Zeltser,
{\it Phys. Lett.} A 192:131 (1994);
A.~S.~Zeltser, A.~E.~Filippov,
{\it Jurn. Exp. Theor. Phys.} 106:1117 (1994) (in Russian).

\bibitem{Malbouisson98}
A.~P.~C.~Malbouisson, F.~S.~Nogueira, N.~F.~Svaiter, {\it Europhys. Lett.}
41:547 (1998).

\bibitem{Coleman73} S.~Coleman, E.~Weinberg, {\it Phys. Rev.} D
7:1988 (1973).

\bibitem{Kang74} J.~S.~Kang, {\it Phys. Rev.} D 10:3455 (1974).

\bibitem{Lawrie82} I.~D.~Lawrie, {\it Nucl. Phys.} B 200:1 (1982).

\bibitem{Hikami79} S.~Hikami, {\it Progr. Theor. Phys.} 26:226 (1979).

\bibitem{Lovesey80} S.~W.~Lovesey, {\it Z. Physik B Condensed Matter}
40:117 (1980)

\bibitem{Dasgupta13} C.~Dasgupta, B.~I.~Halperin, {\it Phys. Rev. Lett.}
47:1556 (1981).

\bibitem{Bartolomew83} J.~Bartholomew, {\it Phys. Rev.} B 28:5378
(1983).

\bibitem{Tonchev81} N.~C.~Tonchev, D.~I.~Uzunov, {\it J. Phys.} A
14:521 (1981).

\bibitem{Boyanovsky82} D.~Boyanovsky, J.~L.~Cardy, {\it Phys. Rev.} B
25:7058 (1982).

\bibitem{Uzunov83}  D.~I.~Uzunov, E.~R.~Korutcheva, Y.~T.~Millev,  {\it J.
Phys.} A 16:247 (1983).

\bibitem{Athorne86} C.~Athorne, I.~D.~Lawrie, {\it Nucl. Phys.} B 265:551 (1986)

\bibitem{Busiello86} G.~Busiello, L.~De~Cesare, D.~I.~Uzunov,
{\it Phys. Rev.} B 34:4932 (1986).

\bibitem{Blagoeva90} E.~J.~Blagoeva, G.~Busiello, L.~De~Cesare,
Y.~T.~Millev, I.~Rabuffo, D.~I.~Uzunov, {\it Phys. Rev.} B 42:6124
(1990).

\bibitem{Busiello91} G.~Busiello, L.~De~Cesare,
Y.~T.~Millev, I.~Rabuffo, D.~I.~Uzunov, {\it Phys. Rev.} B 43:1150
(1991).

\bibitem{Ford92} The expression of the $\zeta$-function for the mass
aggrees with C.~Ford, I.~Jack, D.~Jones, {\it Nucl. Phys.} B387:373
(1992).

\bibitem{Lawrie83} I.~D.~Lawrie, C.~Athorne, {\it J. Phys. A: Math. Gen.}
16:L587 (1983).

\bibitem{Bergerhoff96} B.~Bergerhoff, F.~Freire, D.~F.~Litim, S.~Lola,
C.~Wetterich, {\it Phys. Rev.} B 53:5734 (1996).

\bibitem{Tetradis94} N.~Tetradis, C.~Wetterich, {\it Nucl. Phys.} B
422:541 (1994).

\bibitem{Graeter95} M.~Gr\"ater, C.~Wetterich, {\it Phys. Rev. Lett.}
75:378 (1995).

\bibitem{Arnold94}
P.~Arnold, L.~G.~Yaffe, {\it Phys. Rev.} D 49:3003 (1994);
P.~Arnold, L.~G.~Yaffe, {\it Phys. Rev.} D  55:1114
(1997) (erratum).

\bibitem{Kleinert82} H.~Kleinert, {\it Lett. Nouvo Cimento}  35:405
(1982) (we are thankful to Prof. H. Kleinert for attracting our
attention to this reference).

\bibitem{Kiometzis94} M.~Kiometzis, H.~Kleinert, A.~M.~J.~Schakel,
{\it Phys. Rev. Lett.} 73:1975 (1994), and {\it Fortschr. Phys.} 43:697 (1995).

\bibitem{Bray74} A.~J.~Bray, {\it Phys. Rev. Lett.} 32:1413 (1974).

\bibitem{Radzihovsky95} L.~Radzihovsky, {\it Europhys. Lett.} 29:227
(1995).

\bibitem{Herbut96} I.~F.~Herbut, Z.~Te\u{s}anovi\'c, {\it Phys. Rev. Lett.}
76:4588 (1996).

\bibitem{Herbut97} I.~F.~Herbut, {\it J. Phys.} A 30:423 (1997).

\bibitem{17} In fact an extensive two loop calculation have been
already performed earlier by M.~Machacek and M.~Vaughn, {\it Nucl.
Phys.} B222:83 (1983); B236:221 (1984); B249:70 (1985).

\bibitem{Olsson98} P. Olsson, S. Teitel, {\it Phys. Rev. Lett.}
80:1964 (1998).

\bibitem{hall} see e.g. X.~Wen, Y.~Wu, {\it Phys. Rev. Lett.} 70:1501
(1993) and L.~Pryadko, S.~Zhang, {\it Phys. Rev. Lett.} 73:3282 (1994).

\bibitem{Nanda98} K.~K.~Nanda, B.~Kalta, {\it Phys. Rev.} B 57:123 (1998).

\bibitem{deGennes72} P.~G.~de~Gennes, {\it Solid State Commun.} 10:753 (1972).

\bibitem{Halperin74a} B.~I.~Halperin, T.~C.~Lubensky, {\it Solid
State Commun.} 14:997 (1974).

\bibitem{Lubensky78} T.~C.~Lubensky, J.-H.~Chen, {\it Phys. Rev.} B 17:366
(1978).

\bibitem{Kasting80} G.~B.~Kasting, K.~J.~Lushington, C.~W.~Garland,
{\it Phys. Rev.} B 22:321 (1980).

\bibitem{Anisimov90} M.~A.~Anisimov, P.~E.~Cladis, E.~E.~Gorodetskii,
D.~A.~Huse, V.~E.~Podneks, V.~G.~Taratuta, W.~van~Saarloos,
V.~P.~Voronov, {\it Phys. Rev.} A 41:6749 (1990).

\bibitem{Garland94} For the most updated comprehensive review of the
experimental data on effective critical exponents governing
nematic-smectic-A phase transitions see: C.~W.~Garland, G.~Nounesis,
{\it Phys. Rev.} E 49:2964 (1994).

\bibitem{tHooft72}
G.~t'Hooft, M.~Veltman, {\it Nucl. Phys.} B 44:189 (1972).

\bibitem{note1a} The first index is the number of $\Psi$ fields, the
second index is the number of $A$ fields.

\bibitem{note2}
As an example for a determination of the critical exponents values
for models with complicated symmetry by applying the resummation
technique in different RG schemes
see\refnote{\cite{Jug83,Mayer89,Mayer89a,Holovatch92,Janssen95}};
N.~A.~Shpot, {\it Phys.Lett.} A 142:474 (1989);
S.~A.~Antonenko, A.~I.~Sokolov, {\it Phys. Rev.} B 49:15901 (1994);
C.~von~Ferber, Yu.~Holovatch,
{\it Europhys. Lett.} 39:31 (1997),
{\it Phys. Rev.} E 56:6370 (1997);
H.~Kleinert, S.~Thoms, V.~Schulte-Frohlinde, preprint (1997).

\bibitem{Jug83} G.~Jug, {\it Phys. Rev.} B 27:609 (1983).

\bibitem{Mayer89} I.~O.~Mayer, A.~I.~Sokolov, B.~N.~Shalaev,
{\it Ferroelectrics} 95:93 (1989).

\bibitem{Mayer89a} I.~O.~Mayer, {\it J. Phys.} A 22:2815 (1989).

\bibitem{Holovatch92} Yu.~Holovatch, M.~Shpot, {\it J. Stat. Phys.}
66:867 (1989);
Yu.~Holovatch, T.~Yavors'kii, (1998) submitted to
{\it J. Stat. Phys.}

\bibitem{Janssen95} H.~K.~Janssen, K.~Oerding, E.~Sengespeick, {\it J.
Phys. A: Math. Gen.} 28:6073 (1995).

\bibitem{Grinstein76} G.~Grinstein, A.~Luther, {\it Phys. Rev.} B 13:1329 (1976)

\bibitem{29} G.H. Hardy. "Divergent Series," Oxford University,
Oxford (1948).

\bibitem{note3}  Only when one
of the couplings is equal to zero one does obtain a series which is proven
to be asymptotic.

\bibitem{Mitchell86} P.~W.~Mitchell, R.~A.~Cowley,
H.~Yoshizawa, P.~B\"oni, Y.~J.~Uemura, R.~J.~Birgeneau, {\it Phys. Rev.} B
34:4719 (1986).

\bibitem{Birgeneau88} T.~R.~Thurston, C.~J.~Peters,
R.~J.~Birgeneau, P.~M.~Horn, {\it Phys. Rev.} B 37:9559 (1988).

\bibitem{Wang90} J.-S.~Wang, M.~W\"ohlert, H.~M\"uhlenbein,
D.~Chowdhury, {\it Physica} A 166:173 (1990).

\bibitem{Wang94} J.-S.~Wang, W.~Selke, Vl.~S.~Dotsenko,
V.~B.~Andreichenko, {\it Europhys. Lett.} 11:301 (1994);
A.~L.~Talapov, L.~N.~Shchur, {\it Europhys. Lett.} 27:193 (1994).

\bibitem{Holey90} T.~Holey, M.~F\"ahnle, {\it Phys. Rev.} B 41:11709
(1990).

\bibitem{note4} The results of the resummation appear to be quite
insensitive to the choice of $p$.

\bibitem{Chisholm73} J.~S.~R.~Chisholm, {\it Math. Comp.}
27:841 (1973).

\bibitem{Watson74} P.~J.~S.~Watson, {\it J. Phys.} A 7:L167 (1974).

\bibitem{Baker81} G.~A.~Baker~Jr., P.~Graves-Morris. "Pad\'e
approximants," Addison-Wesley Publ. Co., Reading, Mass. (1981).

\bibitem{Parisi} G.~Parisi, {\it in}: "Proceedings of the 1973
Cargr\'ese Summer School," unpublished.  G.~Parisi, {\it J. Stat. Phys.}
23:49 (1980).

\bibitem{Kleinert95}
H.~Kleinert, V.~Schulte-Frohlinde, {\it Phys. Lett.} B 342:284
(1995).

\bibitem{note5} For this  model the system of fixed point equations is
degenerated at the one-loop level, resulting in particular in the
$\sqrt{\varepsilon}$-expansion for the critical exponents:
D.~E.~Khmelnitskii, {\it Zh. Eksp. Theor. Fiz.} 68:1960
(1975); T.~C.~Lubensky, {\it Phys. Rev.} B 11:3573 (1975);
\refnote{\cite{Grinstein76}}.

\bibitem{note6} It corresponds to a $n=0$ fixed point of the $O(n)$
symmetrical model, described by the de Gennes limit of the self
avoiding walk problem.

\bibitem{note7} The results are given for the value of the parameter
$p=0$ in the Borel-Leroy image.

\bibitem{Holovatch93} Yu.~Holovatch, Preprint, C.E.Saclay, {\it Service
de Physique Theorique}; S Ph T / 92 - 123); Yu.~Holovatch, {\it Int. J. Mod.
Phys.} A 8:5329 (1993).

\bibitem{note8} The last possibility has been chosen by
G.~Parisi\refnote{\cite{Parisi}} in order to restore the
presence of a stable solution for
the fixed point in the two-loop approximation.

\bibitem{49b} The series in (\ref{20}) appears to be an alternating
one and this scheme can be applied without any difficulties.

\bibitem{Abramovitz64} M.~Abramowitz, A.~I.~Stegun, (editors)
"Handbook of Mathematical Functions with Formulas, Graphs and
Mathematical Tables," National Bureau of Standards (1964).

\bibitem{note9}  In this case the principal value of the integral
(\ref{19}) could be taken, but generally speaking it is preferable to
avoid such situations (see\refnote{\cite{Baker78}} as well).

\bibitem{note10} We take $\ell_0=1$.

\bibitem{note11} The value of $\eta$ has been found by the scaling law:
$\eta=2-\gamma/\nu$.

\bibitem{DOPRL} V. Dohm, {\it Phys. Rev. Lett.} 53:1379 (1984).

\bibitem{DOGARR} V. Dohm, {\it in}: "Application of field
theory to statistical mechanics," L. Garrido Ed. p 263,
Berlin, Heidelberg, New York, Tokyo,  Springer (1985).

\bibitem{DOZEIT} V. Dohm, {\it Z. Physik Condensed Matter} 60:61 (1985).

\bibitem{SINGSAAS} A. Singsaas and G. Ahlers, {\it Phys. Rev.} B 30:5103 (1984).

\bibitem{ANNETT} J. F. Annett, S. R. Renn, {\it Phys. Rev.} B 38:4660 (1988).

\bibitem{Riedel74} E. K. Riedel, F. J. Wegner, {\it Phys. Rev.} B
9:294 (1974).

\bibitem{note12} In fact we have solved the flow equations (\ref{37}),
(\ref{38}) starting near the unstable fixed points, for the initial
value of the flow parameter we have taken $\ell=1$. The use of different
initial values ($u(1)$ and $f(1)$ on the seperatrix) would amount to rescale
the flow parameter.

\bibitem{Lawrie97} I.~D.~Lawrie, {\it Phys. Rev. Lett.}  79:131
(1997); see also K.~K.~Nanda, B.~Kalta, {\it Phys. Rev.} B 57:123
(1998).

\bibitem{note12a} Because they are double series in two coupling
constants we have made use of representing them in the form of resolvent
series which has enabled us then to pass to the Pad\'e analysis.

\end{numbibliography}
\end{document}